\documentclass[final]{IEEEtran}
\usepackage{array}
\newcolumntype{P}[1]{>{\centering\arraybackslash}p{#1}}
\newcolumntype{M}[1]{>{\centering\arraybackslash}m{#1}}
\usepackage{amsthm,amssymb,amsmath,graphicx,multirow,color,amsfonts}%
\usepackage[update,prepend]{epstopdf}
\usepackage[latin1]{inputenc}
\usepackage{tikz}
\usepackage{bbm} 
\usepackage{pdfpages}
\usepackage{subfig}
\usepackage{comment}
\usepackage{makecell}
\usepackage{multicol}

\usepackage{cite}
\usepackage[justification=centering]{caption}
\usepackage{textcomp}
\usepackage{psfrag}
\usepackage{arydshln}
\usepackage{url}
\usepackage{soul}
\usepackage{graphicx,color}
\usepackage[nolist]{acronym}
\usepackage{algorithm,algorithmic} 

\usepackage{mathtools,lipsum}
\usepackage{cuted}
\usepackage{amsmath}
 
\usepackage{mathrsfs}

\usepackage[capitalise]{cleveref}
\Crefname{equation}{Eq.\!}{Eqs.\!}
\Crefname{figure}{Fig.\!}{Figs.\!}
\Crefname{tabular}{Tab.\!}{Tabs.\!}
\Crefname{section}{Section\!}{Sections.\!}

\def\nb0{{\mathbf{0}}}
\def\nb1{{\mathbf{1}}}

\newtheorem{lemma}{Lemma}

\newtheorem{definition}{Definition}

\newtheorem{theorem}{Theorem}

\begin{document}
\graphicspath{{./Figures/}}
	\begin{acronym}

\acro{5G-NR}{5G New Radio}
\acro{3GPP}{3rd Generation Partnership Project}
\acro{ABS}{aerial base station}
\acro{AC}{address coding}
\acro{ACF}{autocorrelation function}
\acro{ACR}{autocorrelation receiver}
\acro{ADC}{analog-to-digital converter}
\acrodef{aic}[AIC]{Analog-to-Information Converter}     
\acro{AIC}[AIC]{Akaike information criterion}
\acro{aric}[ARIC]{asymmetric restricted isometry constant}
\acro{arip}[ARIP]{asymmetric restricted isometry property}

\acro{ARQ}{Automatic Repeat Request}
\acro{AUB}{asymptotic union bound}
\acrodef{awgn}[AWGN]{Additive White Gaussian Noise}     
\acro{AWGN}{additive white Gaussian noise}

\acro{APSK}[PSK]{asymmetric PSK} 

\acro{waric}[AWRICs]{asymmetric weak restricted isometry constants}
\acro{warip}[AWRIP]{asymmetric weak restricted isometry property}
\acro{BCH}{Bose, Chaudhuri, and Hocquenghem}        
\acro{BCHC}[BCHSC]{BCH based source coding}
\acro{BEP}{bit error probability}
\acro{BFC}{block fading channel}
\acro{BG}[BG]{Bernoulli-Gaussian}
\acro{BGG}{Bernoulli-Generalized Gaussian}
\acro{BPAM}{binary pulse amplitude modulation}
\acro{BPDN}{Basis Pursuit Denoising}
\acro{BPPM}{binary pulse position modulation}
\acro{BPSK}{Binary Phase Shift Keying}
\acro{BPZF}{bandpass zonal filter}
\acro{BSC}{binary symmetric channels}              
\acro{BU}[BU]{Bernoulli-uniform}
\acro{BER}{bit error rate}
\acro{BS}{base station}
\acro{BW}{BandWidth}
\acro{BLLL}{ binary log-linear learning }

\acro{CP}{Cyclic Prefix}
\acrodef{cdf}[CDF]{cumulative distribution function}   
\acro{CDF}{Cumulative Distribution Function}
\acrodef{c.d.f.}[CDF]{cumulative distribution function}
\acro{CCDF}{complementary cumulative distribution function}
\acrodef{ccdf}[CCDF]{complementary CDF}               
\acrodef{c.c.d.f.}[CCDF]{complementary cumulative distribution function}
\acro{CD}{cooperative diversity}

\acro{CDMA}{Code Division Multiple Access}
\acro{ch.f.}{characteristic function}
\acro{CIR}{channel impulse response}
\acro{cosamp}[CoSaMP]{compressive sampling matching pursuit}
\acro{CR}{cognitive radio}
\acro{cs}[CS]{compressed sensing}                   
\acrodef{cscapital}[CS]{Compressed sensing} 
\acrodef{CS}[CS]{compressed sensing}
\acro{CSI}{channel state information}
\acro{CCSDS}{consultative committee for space data systems}
\acro{CC}{convolutional coding}
\acro{Covid19}[COVID-19]{Coronavirus disease}

\acro{DAA}{detect and avoid}
\acro{DAB}{digital audio broadcasting}
\acro{DCT}{discrete cosine transform}
\acro{dft}[DFT]{discrete Fourier transform}
\acro{DR}{distortion-rate}
\acro{DS}{direct sequence}
\acro{DS-SS}{direct-sequence spread-spectrum}
\acro{DTR}{differential transmitted-reference}
\acro{DVB-H}{digital video broadcasting\,--\,handheld}
\acro{DVB-T}{digital video broadcasting\,--\,terrestrial}
\acro{DL}{DownLink}
\acro{DSSS}{Direct Sequence Spread Spectrum}
\acro{DFT-s-OFDM}{Discrete Fourier Transform-spread-Orthogonal Frequency Division Multiplexing}
\acro{DAS}{Distributed Antenna System}
\acro{DNA}{DeoxyriboNucleic Acid}

\acro{EC}{European Commission}
\acro{EED}[EED]{exact eigenvalues distribution}
\acro{EIRP}{Equivalent Isotropically Radiated Power}
\acro{ELP}{equivalent low-pass}
\acro{eMBB}{Enhanced Mobile Broadband}
\acro{EMF}{ElectroMagnetic Field}
\acro{EU}{European union}
\acro{EI}{Exposure Index}
\acro{eICIC}{enhanced Inter-Cell Interference Coordination}

\acro{FC}[FC]{fusion center}
\acro{FCC}{Federal Communications Commission}
\acro{FEC}{forward error correction}
\acro{FFT}{fast Fourier transform}
\acro{FH}{frequency-hopping}
\acro{FH-SS}{frequency-hopping spread-spectrum}
\acrodef{FS}{Frame synchronization}
\acro{FSsmall}[FS]{frame synchronization}  
\acro{FDMA}{Frequency Division Multiple Access}

\acro{GA}{Gaussian approximation}
\acro{GF}{Galois field }
\acro{GG}{Generalized-Gaussian}
\acro{GIC}[GIC]{generalized information criterion}
\acro{GLRT}{generalized likelihood ratio test}
\acro{GPS}{Global Positioning System}
\acro{GMSK}{Gaussian Minimum Shift Keying}
\acro{GSMA}{Global System for Mobile communications Association}
\acro{GS}{ground station}
\acro{GMG}{ Grid-connected MicroGeneration}

\acro{HAP}{high altitude platform}
\acro{HetNet}{Heterogeneous network}

\acro{IDR}{information distortion-rate}
\acro{IFFT}{inverse fast Fourier transform}
\acro{iht}[IHT]{iterative hard thresholding}
\acro{i.i.d.}{independent, identically distributed}
\acro{IoT}{Internet of Things}                      
\acro{IR}{impulse radio}
\acro{lric}[LRIC]{lower restricted isometry constant}
\acro{lrict}[LRICt]{lower restricted isometry constant threshold}
\acro{ISI}{intersymbol interference}
\acro{ITU}{International Telecommunication Union}
\acro{ICNIRP}{International Commission on Non-Ionizing Radiation Protection}
\acro{IEEE}{Institute of Electrical and Electronics Engineers}
\acro{ICES}{IEEE international committee on electromagnetic safety}
\acro{IEC}{International Electrotechnical Commission}
\acro{IARC}{International Agency on Research on Cancer}
\acro{IS-95}{Interim Standard 95}

\acro{KPI}{Key Performance Indicator}

\acro{LEO}{low earth orbit}
\acro{LF}{likelihood function}
\acro{LLF}{log-likelihood function}
\acro{LLR}{log-likelihood ratio}
\acro{LLRT}{log-likelihood ratio test}
\acro{LoS}{Line-of-Sight}
\acro{LRT}{likelihood ratio test}
\acro{wlric}[LWRIC]{lower weak restricted isometry constant}
\acro{wlrict}[LWRICt]{LWRIC threshold}
\acro{LPWAN}{Low Power Wide Area Network}
\acro{LoRaWAN}{Low power long Range Wide Area Network}
\acro{NLoS}{Non-Line-of-Sight}
\acro{LiFi}[Li-Fi]{light-fidelity}
 \acro{LED}{light emitting diode}
 \acro{LABS}{LoS transmission with each ABS}
 \acro{NLABS}{NLoS transmission with each ABS}

\acro{MB}{multiband}
\acro{MC}{macro cell}
\acro{MDS}{mixed distributed source}
\acro{MF}{matched filter}
\acro{m.g.f.}{moment generating function}
\acro{MI}{mutual information}
\acro{MIMO}{Multiple-Input Multiple-Output}
\acro{MISO}{multiple-input single-output}
\acrodef{maxs}[MJSO]{maximum joint support cardinality}                       
\acro{ML}[ML]{maximum likelihood}
\acro{MMSE}{minimum mean-square error}
\acro{MMV}{multiple measurement vectors}
\acrodef{MOS}{model order selection}
\acro{M-PSK}[${M}$-PSK]{$M$-ary phase shift keying}                       
\acro{M-APSK}[${M}$-PSK]{$M$-ary asymmetric PSK} 
\acro{MP}{ multi-period}
\acro{MINLP}{mixed integer non-linear programming}

\acro{M-QAM}[$M$-QAM]{$M$-ary quadrature amplitude modulation}
\acro{MRC}{maximal ratio combiner}                  
\acro{maxs}[MSO]{maximum sparsity order}                                      
\acro{M2M}{Machine-to-Machine}                                                
\acro{MUI}{multi-user interference}
\acro{mMTC}{massive Machine Type Communications}      
\acro{mm-Wave}{millimeter-wave}
\acro{MP}{mobile phone}
\acro{MPE}{maximum permissible exposure}
\acro{MAC}{media access control}
\acro{NB}{narrowband}
\acro{NBI}{narrowband interference}
\acro{NLA}{nonlinear sparse approximation}
\acro{NLOS}{Non-Line of Sight}
\acro{NTIA}{National Telecommunications and Information Administration}
\acro{NTP}{National Toxicology Program}
\acro{NHS}{National Health Service}

\acro{LOS}{Line of Sight}

\acro{OC}{optimum combining}                             
\acro{OC}{optimum combining}
\acro{ODE}{operational distortion-energy}
\acro{ODR}{operational distortion-rate}
\acro{OFDM}{Orthogonal Frequency-Division Multiplexing}
\acro{omp}[OMP]{orthogonal matching pursuit}
\acro{OSMP}[OSMP]{orthogonal subspace matching pursuit}
\acro{OQAM}{offset quadrature amplitude modulation}
\acro{OQPSK}{offset QPSK}
\acro{OFDMA}{Orthogonal Frequency-division Multiple Access}
\acro{OPEX}{Operating Expenditures}
\acro{OQPSK/PM}{OQPSK with phase modulation}

\acro{PAM}{pulse amplitude modulation}
\acro{PAR}{peak-to-average ratio}
\acrodef{pdf}[PDF]{probability density function}                      
\acro{PDF}{probability density function}
\acrodef{p.d.f.}[PDF]{probability distribution function}
\acro{PDP}{power dispersion profile}
\acro{PMF}{probability mass function}                             
\acrodef{p.m.f.}[PMF]{probability mass function}
\acro{PN}{pseudo-noise}
\acro{PPM}{pulse position modulation}
\acro{PRake}{Partial Rake}
\acro{PSD}{power spectral density}
\acro{PSEP}{pairwise synchronization error probability}
\acro{PSK}{phase shift keying}
\acro{PD}{power density}
\acro{8-PSK}[$8$-PSK]{$8$-phase shift keying}
\acro{PPP}{Poisson point process}
\acro{PCP}{Poisson cluster process}
 
\acro{FSK}{Frequency Shift Keying}

\acro{QAM}{Quadrature Amplitude Modulation}
\acro{QPSK}{Quadrature Phase Shift Keying}
\acro{OQPSK/PM}{OQPSK with phase modulator }

\acro{RD}[RD]{raw data}
\acro{RDL}{"random data limit"}
\acro{ric}[RIC]{restricted isometry constant}
\acro{rict}[RICt]{restricted isometry constant threshold}
\acro{rip}[RIP]{restricted isometry property}
\acro{ROC}{receiver operating characteristic}
\acro{rq}[RQ]{Raleigh quotient}
\acro{RS}[RS]{Reed-Solomon}
\acro{RSC}[RSSC]{RS based source coding}
\acro{r.v.}{random variable}                               
\acro{R.V.}{random vector}
\acro{RMS}{root mean square}
\acro{RFR}{radiofrequency radiation}
\acro{RIS}{Reconfigurable Intelligent Surface}
\acro{RNA}{RiboNucleic Acid}
\acro{RRM}{Radio Resource Management}
\acro{RUE}{reference user equipments}
\acro{RAT}{radio access technology}
\acro{RB}{resource block}

\acro{SA}[SA-Music]{subspace-augmented MUSIC with OSMP}
\acro{SC}{small cell}
\acro{SCBSES}[SCBSES]{Source Compression Based Syndrome Encoding Scheme}
\acro{SCM}{sample covariance matrix}
\acro{SEP}{symbol error probability}
\acro{SG}[SG]{sparse-land Gaussian model}
\acro{SIMO}{single-input multiple-output}
\acro{SINR}{signal-to-interference plus noise ratio}
\acro{SIR}{signal-to-interference ratio}
\acro{SISO}{Single-Input Single-Output}
\acro{SMV}{single measurement vector}
\acro{SNR}[\textrm{SNR}]{signal-to-noise ratio} 
\acro{sp}[SP]{subspace pursuit}
\acro{SS}{spread spectrum}
\acro{SW}{sync word}
\acro{SAR}{specific absorption rate}
\acro{SSB}{synchronization signal block}
\acro{SR}{shrink and realign}

\acro{tUAV}{tethered Unmanned Aerial Vehicle}
\acro{TBS}{terrestrial base station}

\acro{uUAV}{untethered Unmanned Aerial Vehicle}
\acro{PDF}{probability density functions}

\acro{PL}{path-loss}

\acro{TH}{time-hopping}
\acro{ToA}{time-of-arrival}
\acro{TR}{transmitted-reference}
\acro{TW}{Tracy-Widom}
\acro{TWDT}{TW Distribution Tail}
\acro{TCM}{trellis coded modulation}
\acro{TDD}{Time-Division Duplexing}
\acro{TDMA}{Time Division Multiple Access}
\acro{Tx}{average transmit}

\acro{UAV}{Unmanned Aerial Vehicle}
\acro{uric}[URIC]{upper restricted isometry constant}
\acro{urict}[URICt]{upper restricted isometry constant threshold}
\acro{UWB}{ultrawide band}
\acro{UWBcap}[UWB]{Ultrawide band}   
\acro{URLLC}{Ultra Reliable Low Latency Communications}
         
\acro{wuric}[UWRIC]{upper weak restricted isometry constant}
\acro{wurict}[UWRICt]{UWRIC threshold}                
\acro{UE}{User Equipment}
\acro{UL}{UpLink}

\acro{WiM}[WiM]{weigh-in-motion}
\acro{WLAN}{wireless local area network}
\acro{wm}[WM]{Wishart matrix}                               
\acroplural{wm}[WM]{Wishart matrices}
\acro{WMAN}{wireless metropolitan area network}
\acro{WPAN}{wireless personal area network}
\acro{wric}[WRIC]{weak restricted isometry constant}
\acro{wrict}[WRICt]{weak restricted isometry constant thresholds}
\acro{wrip}[WRIP]{weak restricted isometry property}
\acro{WSN}{wireless sensor network}                        
\acro{WSS}{Wide-Sense Stationary}
\acro{WHO}{World Health Organization}
\acro{Wi-Fi}{Wireless Fidelity}

\acro{sss}[SpaSoSEnc]{sparse source syndrome encoding}

\acro{VLC}{Visible Light Communication}
\acro{VPN}{Virtual Private Network} 
\acro{RF}{Radio Frequency}
\acro{FSO}{Free Space Optics}
\acro{IoST}{Internet of Space Things}

\acro{GSM}{Global System for Mobile Communications}
\acro{2G}{Second-generation cellular network}
\acro{3G}{Third-generation cellular network}
\acro{4G}{Fourth-generation cellular network}
\acro{5G}{Fifth-generation cellular network}	
\acro{gNB}{next-generation Node-B Base Station}
\acro{NR}{New Radio}
\acro{UMTS}{Universal Mobile Telecommunications Service}
\acro{LTE}{Long Term Evolution}

\acro{QoS}{Quality of Service}
\end{acronym}
	
\newcommand{\SAR} {\mathrm{SAR}}
\newcommand{\WBSAR} {\mathrm{SAR}_{\mathsf{WB}}}
\newcommand{\gSAR} {\mathrm{SAR}_{10\si{\gram}}}
\newcommand{\Sab} {S_{\mathsf{ab}}}
\newcommand{\Eavg} {E_{\mathsf{avg}}}
\newcommand{\ft}{f_{\textsf{th}}}
\newcommand{\alphatf}{\alpha_{24}}

\title{
Coverage Analysis of Large-Scale HAPS \\
Networks Using Directional Beams 
}
\author{
Zhengying Lou, Baha Eddine Youcef Belmekki,~\IEEEmembership{Senior Member,~IEEE,} and Mohamed-Slim Alouini, {\em Fellow, IEEE}

\thanks{Z.~Lou and M.~-S.~Alouini are with King Abdullah University of Science and Technology (KAUST), CEMSE division, Thuwal 23955-6900, Saudi Arabia (e-mail: zhengying.lou@kaust.edu.sa; slim.alouini@kaust.edu.sa).
B.~E.~Y.~Belmekki with the School of Engineering and Physical Sciences, Heriot-Watt University, Edinburgh EH14 4AS, United Kingdom (e-mail: b.belmekki@hw.ac.uk).}
\vspace{-8mm}
}
\maketitle

\begin{abstract}
High-altitude platform stations (HAPS) are pivotal in next-generation wireless networks for reducing core network burdens and enabling cost-effective communication.
In this article, we propose a spherical stochastic geometry-based analytical framework for the coverage performance evaluation of HAPS networks. Considering the significant influence of directional antenna gain on interference evaluation, we analyze coverage performance under a general channel model that accommodates various beam patterns. Analytical expressions of the uplink and downlink coverage probabilities in cellular and cell-free networks are provided respectively and their accuracy are verified by Monte Carlo simulation. Furthermore, the influences of network-level and physical-level parameters on coverage probability are studied. Finally, several factors that align with the analytical framework of this article are discussed. 
\end{abstract}


\vspace{-0.2cm}
\section{Introduction}
\vspace{-0.2cm}
\subsection{Motivation}
In the next generation of wireless communication networks, particularly in non-terrestrial networks, high-altitude platform stations (HAPS) are expected to play a pivotal role in reducing the load on core communication networks while facilitating cost-effective connectivity \cite{10417095, 10474118, 10279432,belmekki02024noma}. HAPS offer widespread wireless connectivity in regions where fiber installation is expensive or challenging, such as remote and sparsely populated rural areas \cite{10192297, belmekki2022unleashing}. Moreover, HAPS offer a range of flexible deployment options that make them indispensable for various scenarios, including emergency response and disaster recovery efforts \cite{matracia2022post}. In recent years, the Google Loon project alone has launched a total of more than $600$ HAPS to provide network coverage \cite{serrano2021balloons}. The deployment of large-scale HAPS networks brings opportunities for wireless network connectivity, but it also presents challenges for evaluating network quality.

\par

Most existing studies focus on the behavior and performance of an individual HAPS \cite{tian2021stochastic,xing2021high,swaminathan2021haps}, which corresponds to small-scale HAPS systems. 
However, evaluating the overall performance of the network is more critical for large-scale HAPS networks. 
Moreover, co-channel interference among HAPS significantly affects large-scale HAPS networks. 
Performance evaluations must account for both serving and interference links, causing the number of links to increase quadratically with the number of HAPS. Consequently, the computational complexity of performance evaluation through numerical simulations grows rapidly with network scale. \cite{xing2021high,shibata2020system}. Motivated by the need for a low-complexity and accurate evaluation method for large-scale HAPS networks, this paper proposes an interference-considerate analytical framework for both uplink and downlink.

\vspace{-0.2cm}

\subsection{Related Works}

\par
Stochastic geometry (SG) is a powerful mathematical tool to study large-scale networks, and it is also one of the few methods capable of interference analysis \cite{wang2022ultra}. Furthermore, the computational complexity of the analytical expressions derived under the SG framework does not increase with the number of devices \cite{wang2024ultra}.
Several studies have attempted to analyze the performance of large-scale HAPS networks within the SG framework \cite{wei2023spectrum,xu2023space,gao2019spectrum,lou2023haps,huang2023system}. Among them, authors in \cite{lou2023haps,huang2023system} modeled HAPS as a spherical binomial point process (BPP) and analyzed the network's performance in terms of coverage, channel capacity, and energy efficiency with a case study approach. Furthermore, authors in \cite{wei2023spectrum,xu2023space,gao2019spectrum} established comprehensive analytical frameworks for HAPS networks and derived analytical expressions for the coverage probability. The spherical Poisson point process (PPP) is adopted in \cite{wei2023spectrum,xu2023space,gao2019spectrum} to model HAPS. The main difference between the two distributions is that the number of HAPS in a given region is a deterministic value under the BPP model, while it is a random variable under the PPP model. Compared to BPP, PPP is more suitable for modeling networks in non-closed areas \cite{wang2022stochastic}. Therefore, PPP is adopted in this article, and we will demonstrate through numerical results that the impact of the two modeling approaches on performance analysis is negligible.

\par

 As for channel modeling, the authors in \cite{loo1985statistical} provided channel models of air-to-ground links for high-altitude equipment, and concluded that the multi-path propagation and shadowing effects are primary factors influencing communication.  Furthermore, the direct/serving links between HAPS and users are typically line-of-sight (LoS). While classic Rician models offer simple expressions for analyzing LoS links, shadowed rician (SR) fading can further incorporate shadowing effects for more accurate characterization\cite{na2021performance,hu2024performance}. Moreover, interference links usually cannot maintain a stable LoS and are more susceptible to reflections, diffraction, and scattering.  Consequently, the Nakagami-m model is better suited to characterize the statistical properties of interference signals \cite{tian2021stochastic,hu2024performance}.

\par

 Several existing studies investigate the performance of large-scale wireless networks employing directional beam transmission.  The flat-top antenna pattern is the most commonly used model due to its simple mathematical formulation, which reduces computational complexity \cite{di2015stochastic, thornburg2016performance, dong2024stochastic}. However, this antenna pattern fails to capture the roll-off characteristics of actual antenna patterns, leading to deviations in performance evaluation \cite{yu2017coverage}. To better represent actual antenna properties, \cite{liu2024space} introduces an antenna pattern where parameters are linked to frequency and dimensions through physical formulas. Meanwhile, authors in \cite{dabiri2020analytical} employ a sectorized-cosine directional antenna pattern and consider a bidirectional alignment mechanism in UAV networks. Specifically, the transmitter directs its beam toward the receiver to reduce interference to others, while the receiver orients its reception to filter out interference from surrounding devices.

To sum up, SG is an effective tool for analyzing the performance of large-scale HAPS network. Moreover, employing SR fading for serving links and nakagami-m fading for interference links is consistent with air-ground propagation characteristics.
Additionally, bidirectional beam alignment mechanism is crucial for reducing interference. However, to the best of the authors' knowledge, no studies have investigated the uplink and downlink coverage probability of HAPS networks with bidirectional beam alignment. The lack of studies may be attributed to the difficulty of interference analysis in large-scale HAPS network.


\vspace{-0.2cm}

\subsection{Contribution}
 In this paper, we present an analytical framework based on spherical SG for evaluating the performance of large-scale HAPS network. 
Next, the specific contributions and distinctions from existing research \cite{wei2023spectrum,xu2023space,gao2019spectrum,lou2023haps,huang2023system} are outlined as follows:

\begin{itemize}
    \item Unlike existing studies that focus solely on downlink performance analysis in cellular networks, we have significantly broadened the scope of research. Specifically, we derive the analytical expressions of the interference Laplace transform and coverage probability of cellular and cell-free HAPS networks in the uplink and downlink transmission scenarios, respectively. 
    \item This is the first article to introduce a bidirectional beam alignment mechanism into the spherical SG framework, and this is the most comprehensive study on beam models within the SG framework. We provide flexibility for antenna patterns, and investigate their impact on coverage probability.
    \item We analyze the impacts of network-level parameters (e.g., coverage thresholds and device density) and physical-level parameters (e.g., beam antenna patterns and half-power beamwidth) on coverage probability. The performances of cellular and cell-free networks are compared. Additionally, we provide insights into potential open issues,  the analytical framework evaluated in this paper serves as a solid foundation and powerful tool for future research of HAPS network.
\end{itemize}

The rest of this paper is organized as follows. The system model of HAPS network is presented in Section \ref{SM}. Analytical expressions of uplink and downlink coverage probabilities are derived in Section \ref{UA} and Section \ref{DA}, respectively. Section \ref{NM} shows the comprehensive numerical results. Section \ref{IPOIRD} discusses potential open issues and future research directions. Section \ref{C} concludes this paper.

\vspace{-0.2cm}

\section{System Model} \label{SM}
\subsection{Distribution Model}
\begin{figure}[t]
	\centering
    \includegraphics[width=0.85\linewidth]{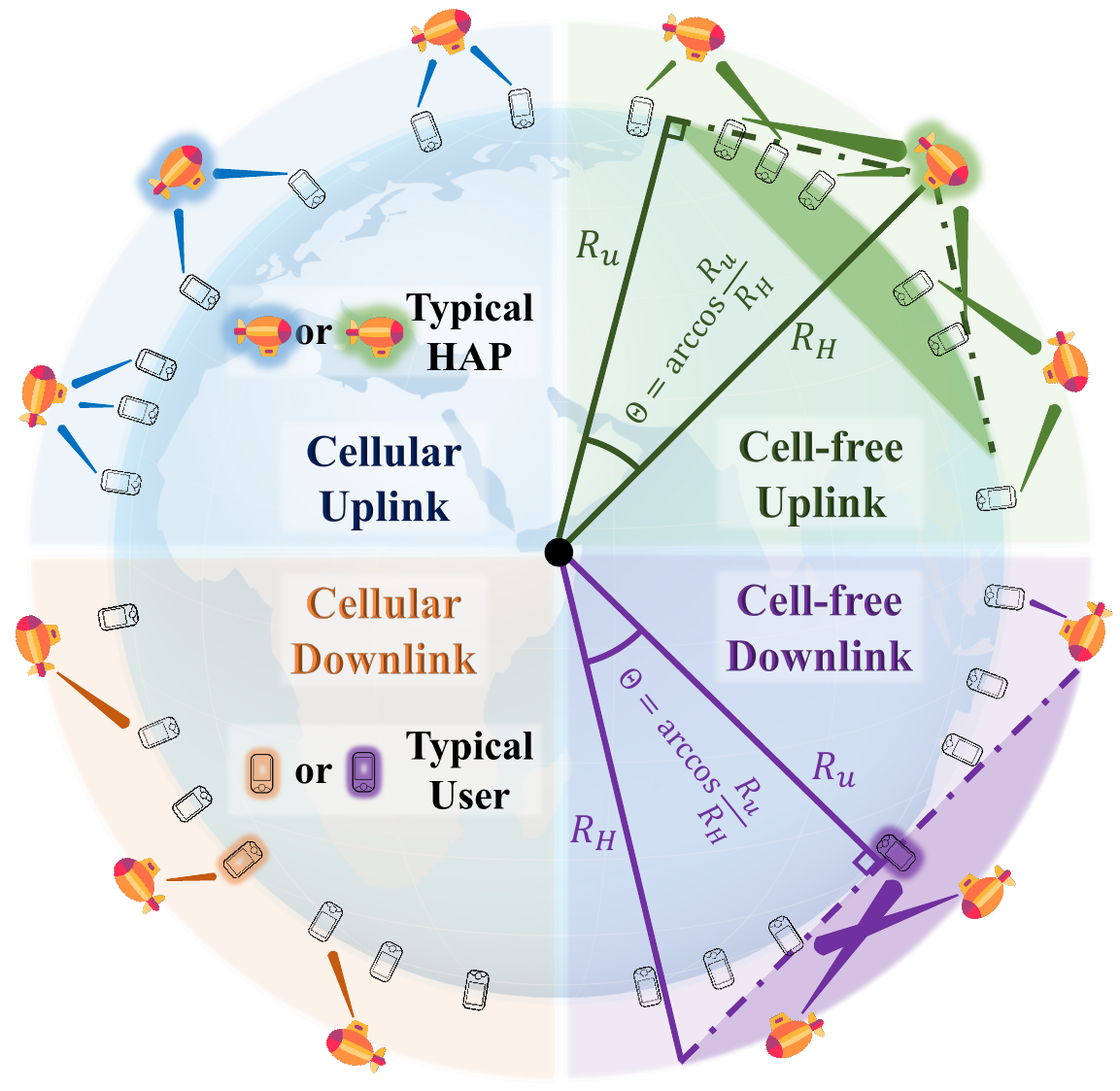}
	\caption{Schematic diagram of cellular and cell-free HAPS networks.}
	\label{fig:Figure1}
	\vspace{-0.5cm}
\end{figure}

We consider a scenario in which multiple HAPS provide services for ground users in a massive air network, as depicted in Fig.~\ref{fig:Figure1}. Users are uniformly distributed over a local region of a sphere with radius $R_u$ and forming a PPP with density $\lambda_u$, where $R_u$ is Earth's radius.\footnote{ We denote Earth's radius by  $R_u$ instead of the commonly used  $R_\oplus$, thereby also indicating users' radius. Although  Earth is oblate, it is typically approximated as a sphere with its average radius within SG framework to reduce computational complexity. Since Earth's flattening ratio is only 0.3\%, and this deviation has a negligible impact on performance evaluation \cite{hmamouche2021new}.}
HAPS are distributed over a local region of a sphere with radius $R_H$ above the user distribution region and forming a PPP with density $\lambda_H$. 

\vspace{-0.2cm}
\subsection{Cellular and Cell-Free}
According to different association strategies, the type of network can be further classified into cellular or cell-free, as shown in Fig.~\ref{fig:Figure1}. In cellular networks, the user adopts the strongest average received power association strategy, while in a cell-free network, the user associates with a HAPS within its LoS region. 
The user's LoS region is defined as the area of the sky where HAPS are visible to the user without any obstructions caused by the Earth's curvature, as indicated in the bottom right of Fig.~\ref{fig:Figure1}.  The whole spectrum resource is divided into a set of orthogonal channels, and then evenly distributed among all users. Here we focus on the analysis of co-channel interference in a particular sub-band. 

The associated HAPS provides communication services to the user, while the power from other HAPS within the user's LoS is considered as interference. Due to the Earth's occlusion, the interference power from HAPS in the non-line-of-sight (NLoS) region is weak, thus, it is ignored in the analysis. Note that both in cellular and cell-free networks, one user can receive only one HAPS service, but one HAPS might be associated with multiple users. As mentioned, we assume that there is at least one HAPS in the user's LoS region. Subsequent numerical results also verify that the impact of HAPS availability is negligible. 

\vspace{-0.2cm}
\subsection{Uplink and Downlink Coordinate Systems}

\begin{figure}[t]
	\centering
  \setlength{\abovecaptionskip}{0.2cm}
\includegraphics[width=0.99\linewidth]{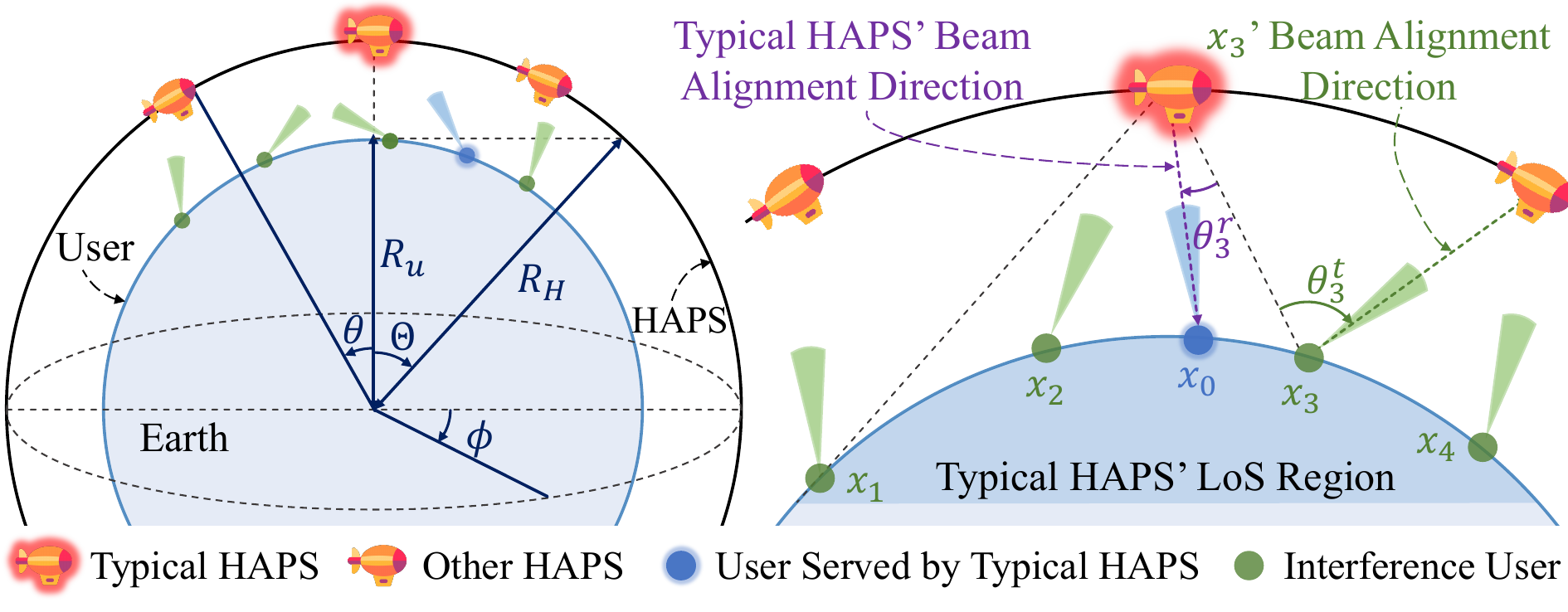}
\caption{Uplink parameter description (cell-free network in the left and cellular network in the right). }
	\label{fig:Figure1_1}
    \vspace{-0.5cm}
\end{figure}

In uplink transmission, and without loss of generality, we focus on the coverage performance of a typical HAPS. We assume that the polar angle is zero at the ray from the Earth center to typical HAPS, and establish a spherical coordinate system. Therefore, the coordinate of typical HAPS is $(R_H,0,0)$. We assume that the locations of users in the LoS range of typical HAPS constitute a set $\mathcal{X}$:
\begin{equation}
\begin{split}
    \mathcal{X} = & \{ x_0(R_u,\phi_0,\theta_0),  x_1(R_u,\phi_1,\theta_1), \dots,\\
    & x_{K-1}(R_u,\phi_{K-1},\theta_{K-1}),  x_K(R_u,\phi_K,\theta_K)\},
\end{split}
\end{equation}
where $x_k(R_u,\phi_k,\theta_k)$ is the coordinate of the $k^{th}$ user, $\phi_k$ and $\theta_k$ denote azimuth angle and polar angle, respectively. The condition $\theta_k \leq \Theta = \arccos({R_u}/{R_H})$ is satisfied for any $0 \leq k \leq K$ to ensure users are in the LoS region. $x_0(R_u,\phi_0,\theta_0)$ is the coordinate of the user being served by the typical HAPS, as indicated in Fig.~\ref{fig:Figure1_1}. Note that due to the Earth's curvature, the altitude of HAPS plays a critical role in determining its LoS region. Specifically, as the altitude increases, the LoS region expands.  Conventionally, HAPS fly in the stratosphere at altitudes of 17-20 km above sea level.

\par
\begin{figure}[t]
	\centering
  \setlength{\abovecaptionskip}{0.2cm}
\includegraphics[width=0.99\linewidth]{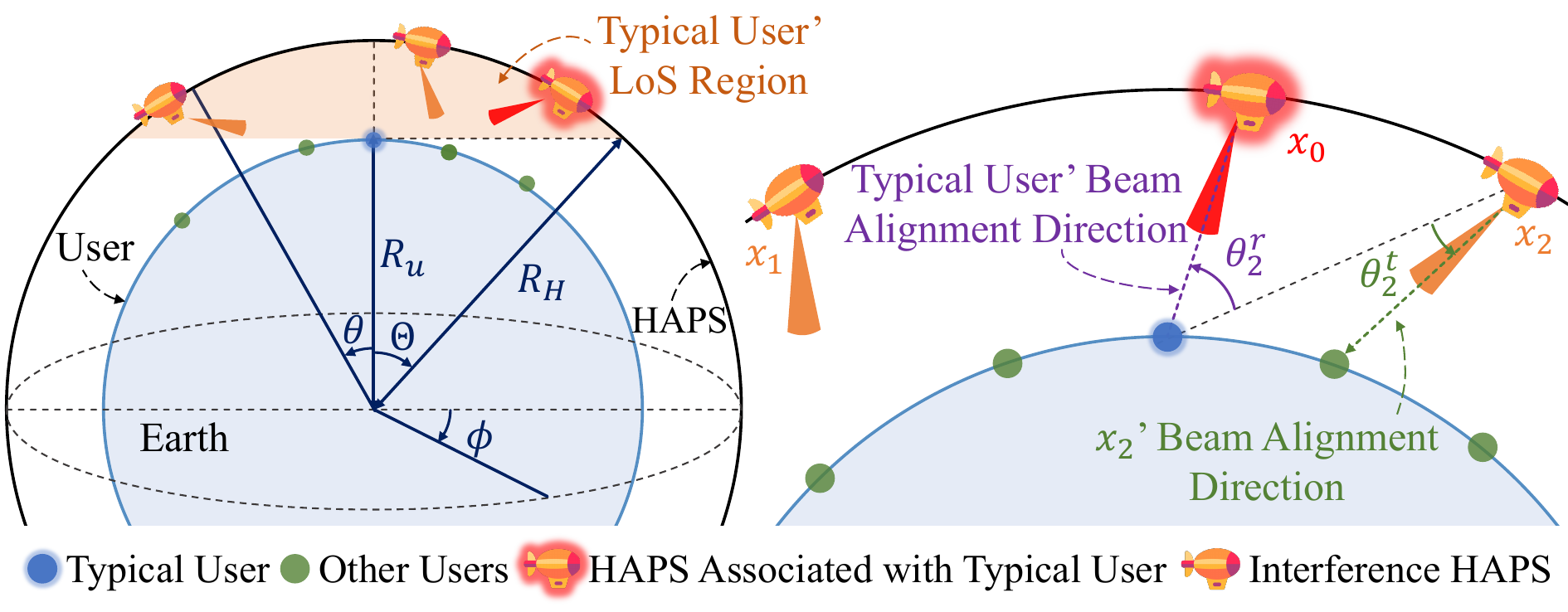}
\caption{Downlink parameter description (cell-free network in the left and cellular network in the right). }
	\label{fig:Figure1_2}
    \vspace{-0.6cm}
\end{figure}

The same method is used to establish a coordinate system for downlink transmission. In this case, the coordinate of the typical user is $(R_u,0,0)$ and $\mathcal{X}$ denotes the positions of HAPS in the typical user's LoS region, as shown in Fig.\ref{fig:Figure1_2}.

\vspace{-0.2cm}
\subsection{Channel Model}
Based on the existing literature \cite{belmekki2022unleashing,swaminathan2021haps}, we consider that the channel model of the HAPS-user link follows the free space fading model experienced with large-scale fading and small-scale fading \cite{na2021performance}.  Moreover, HAPS typically remain quasi-stationary in the stratosphere, making the impact of drift speed negligible, as demonstrated by the world's first 5G HAPS communication trial \cite{10474118}. 
In uplink transmission, the received power of HAPS from the $k^{th}$ user is,
\begin{equation}\label{channel}
    \rho^r_{H,k} = \rho^t_u \, G_u(\theta_k^t) \, G_H(\theta_k^r) \left(\frac{\nu}{4 \pi}\right)^2 \zeta D_k^{-2} W_Q, 
\end{equation}
where $\rho^t_u$ is the transmitted power from the user, $G_u(\theta_k^t)$ and $G_H(\theta_k^r)$ are antenna gains, $\theta_k^t$ and $\theta_k^r$ are angles, $\left(\frac{\nu}{4 \pi}\right)^2 \zeta D_k^{-2}$ denotes the large-scale fading, and $W_Q, \, Q=\{S, I\}$ denotes the small-scale fading. Next, we will discuss these notations in detail.

\begin{table}[ht]
\centering
\setlength{\abovedisplayskip}{2.5pt}
\setlength{\belowdisplayskip}{2.5pt}
\caption{Antenna gains when antenna patterns are adopted.}
\label{table1}
\renewcommand\arraystretch{1}
\begin{tabular}{|cc|}
\hline
\multicolumn{1}{|c|}{Antenna Pattern} &
  Model \\ \hline
\multicolumn{1}{|c|}{Omnidirectional \cite{sun2015synthesizing}} &
  $G(\theta) = G_m$ \\ \hline
\multicolumn{1}{|c|}{Gaussian \cite{gagliardi2012satellite}} &
  $G(\theta) = G_m 2^{-\frac{\theta^2}{ \theta_{\rm 3dB}^2}}$ \\ \hline
\multicolumn{1}{|c|}{Flat-top \cite{balanis2015antenna}} &
  $ G(\theta) = \left\{\begin{matrix} G_m  & \left | \theta \right | \leq  \theta_{\rm 3dB} \\  0    &  { \rm otherwise} \end{matrix}\right.$ \\ \hline
\multicolumn{1}{|c|}{Sinc \cite{yu2017coverage}} &
  $ G(\theta) = G_m \frac{\sin^2\left ( \pi N \theta \right )}{\left ( \pi N \theta \right )^2}$ \\ \hline
\multicolumn{1}{|c|}{Cosine \cite{yu2017coverage}} &
  $ G(\theta) = \left\{\begin{matrix} G_m   \cos^2\left ( \frac{\pi N}{2} \theta \right )  &  \left | \theta \right |   \leq  \frac{1}{N}\\ 0    &  { \rm otherwise} \end{matrix}\right.$ \\ \hline
Parameter &
  \begin{tabular}[c]{@{}c@{}}$G_m$: maximum gain\\ $\theta_{\rm 3dB}$: half-power beamwidth\\ $N$: number of  antenna elements\end{tabular} \\ \hline
\end{tabular}
\vspace{-0.2cm}
\end{table}

\par
\begin{figure}[t]
	\centering
    \includegraphics[width=0.8 \linewidth]{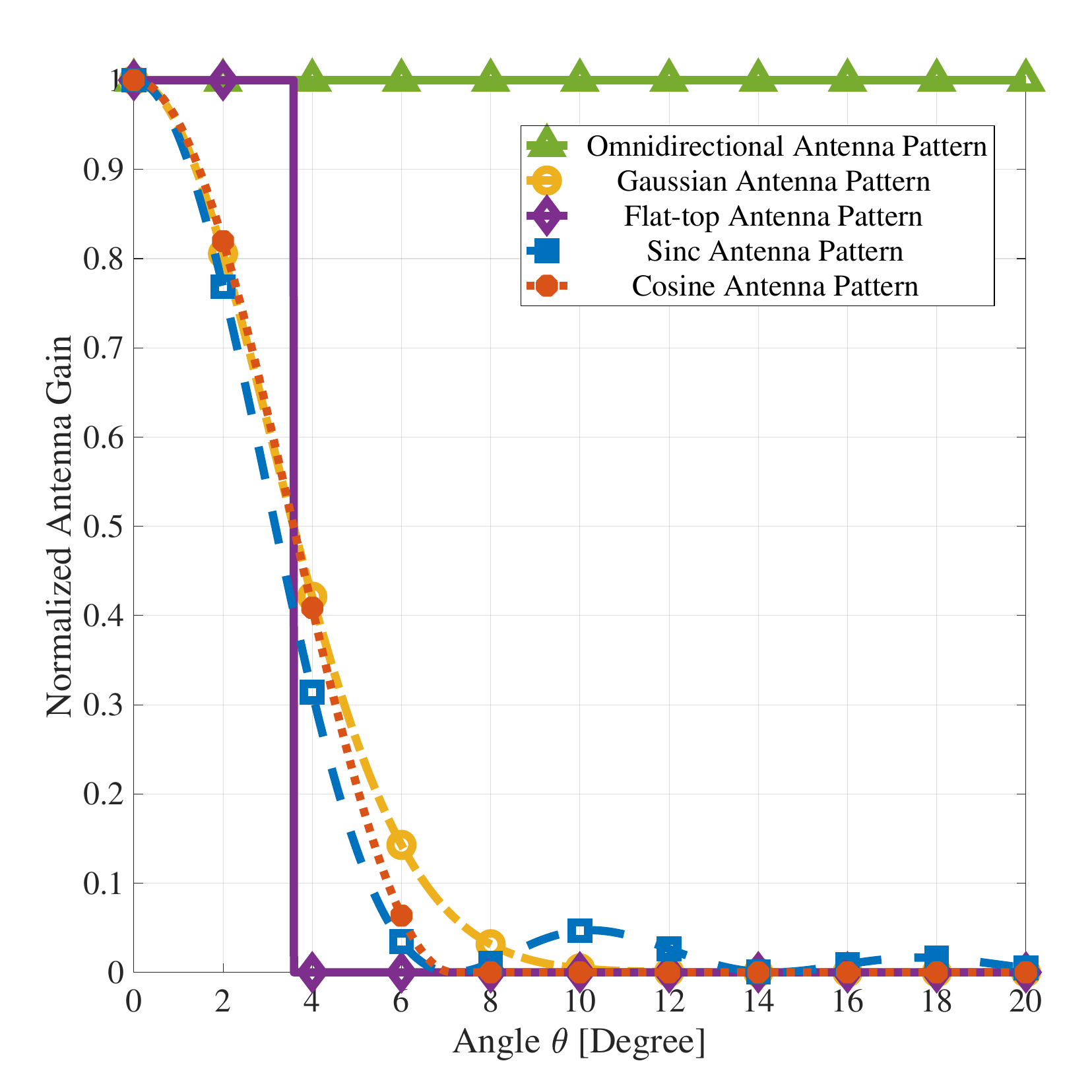}
	\caption{Visualization of five different antenna patterns when $N=8$, $\theta_{\rm 3dB}=3.581^\circ$.}
	\label{fig:FigureAnn}
	\vspace{-0.5cm}
\end{figure}

Here $G(\theta)$ represents the general antenna gain, which can be replaced by specific models when different antenna patterns are adopted. Table~\ref{table1} enumerates five common models, and their visual representations are provided in Fig.\ref{fig:FigureAnn}.

$G_u(\theta_k^t)$ and $G_H(\theta_k^r)$ represent the antenna gain when antenna patterns are applied by the user and HAPS respectively. $\theta_k^t$ is the angle between the $k^{th}$ user's beam alignment direction and the user-HAPS direction. $\theta_k^r$ is the angle between the HAPS' beam alignment direction and the user-HAPS direction. When HAPS is serving the $k^{th}$ user, the user and HAPS' beams are bidirectionally aligned ($\theta_k^t=\theta_k^r=0$), and the total antenna gain reaches its maximum. When HAPS is associated with the $k^{th}$ user but is serving another user, only the user's beam is aligned ($\theta_k^t=0, \theta_k^r\neq0$). Since the communication distance can reach tens or even hundreds of kilometers, we consider the transmitter will not adopt the omnidirectional antenna pattern. 

\par
As for large-scale fading $\left(\frac{\nu}{4 \pi}\right)^2 \zeta D_k^{-2}$, $\nu$ is the wave length of the transmitted signal. $\zeta$ denotes the additional gain during transmission in the medium. In this scenario, $\zeta$ represents the average rain attenuation \cite{talgat2020stochastic}. $D_k$ is the Euclidean distance between the $k^{th}$ user and HAPS. When the distributions of HAPS and user both follow PPPs, $D_k$ is a random variable subject to a specific distribution. Especially, when the user associates with the HAPS, the distribution of $D_k$ is called contact distance distribution \cite{wang2022conditional}. Considering that HAPS and users are distributed on spheres, using the central angle in the spherical coordinate system instead of the Euclidean distance can make the result more tractable. 
\begin{definition}[Central Angle]
The central angle between a HAPS and a user is an angle whose vertex is the Earth center and whose sides are the line from the Earth center to the user and the line from the Earth center to the HAPS.
\end{definition}
Considering the above definition, the relationship between $D_k$ and central angle $\theta_k$ can be established by law of cosines,
\begin{equation}
    \cos\theta_k = \frac{R_u^2+R_H^2-D_k^2}{2R_u R_H}.
\end{equation}

\par
According to \cite{tian2021stochastic}, we assume that the small-scale fading of interference signal $W_I$ follows Nakagami-$m$ fading, and that of the non-interference signal $W_S$ follows shadowed Rician fading. The probability density functions (PDF) of Nakagami-$m$ fading is given by \cite{nakagami}: 
\begin{equation}
    f_{W_I}\left( g \right) = \frac{{m^m g^{m - 1}}}{{\Gamma \left( m \right)}}{e^{ - m \, g}},
\end{equation}
where $\Gamma \left( m \right) = \int_0^\infty  {z^{m - 1}{e^{ - z}}dz}$ is the Gamma function, $m$ is the scale parameter of Nakagami-$m$ fading. The cumulative distribution function (CDF) of shadowed Rician fading is given as by \cite{na2021performance}
\begin{IEEEeqnarray}{RCL}
     F_{W_S} (g) & = &\left( \frac{2 b_0 n}{2 b_0 n + \Omega} \right)^n  \notag \\
     & \times& \sum_{z=0}^{\infty} \frac{(n)_z}{z! \Gamma(z+1)} \left( \frac{\Omega}{2 b_0 n + \Omega} \right)^z \Gamma\left(z+1, \frac{g}{2b_0} \right), \IEEEeqnarraynumspace
\end{IEEEeqnarray}
where $(n)_z$ is the Pochhammer symbol, $n$, $b_0$ and $\Omega$ are parameters of the shadowed Rician fading. $\Gamma( \cdot )$ and $\Xi(\cdot, \cdot )$ denote the gamma function and lower incomplete gamma function, respectively. 

\par
Finally, we assume downlink transmission experience the same channel fading as uplink transmission. Therefore, the received power of the user is given by
\begin{equation}
    \rho^r_{u,k} = \rho^t_H \, G_u(\theta_k^r) \, G_H(\theta_k^t) \left(\frac{\nu}{4 \pi}\right)^2 \zeta D_k^{-2} W_Q, 
\end{equation}
where $\rho^t_H$ is the transmitted power from the HAPS.

\subsection{Metric Definition}
This subsection gives the definition of coverage probability in uplink transmission. The definition of coverage probability in downlink transmission is similar to that of uplink transmission, thus omitted here. 

\par
From the above descriptions, the received power from the user at $x_0$ associated by the typical HAPS is given as
\begin{equation}
    \rho^r_{H,0} = \rho^t_u \, G_u(0) \, G_H(0) \left(\frac{\nu}{4 \pi}\right)^2 \zeta D_0^{-2} W_S.
\end{equation}
The total interference received by the typical HAPS is the sum of the received power from $K$ other users:
\begin{IEEEeqnarray}{RCL}
    I & =& \sum_{x_i \in \mathcal{X} \backslash x_0} \rho_{H,x_i}^r  \notag \\
    & =& \sum_{k=1}^{K} \rho^t_u \, G_u(\theta_k^t) \, G_H(\theta_k^r) \left(\frac{\nu}{4 \pi}\right)^2 \zeta D_k^{-2} W_I.\IEEEeqnarraynumspace
\end{IEEEeqnarray}
The coverage probability is defined as the probability that the received signal-to-interference plus noise ratio (SINR) at the typical HAPS is greater than a coverage threshold $\gamma_H$, such as
\begin{IEEEeqnarray}{RCL}
   & & \!\!\!P_u^{C,Y} (\gamma_H) =  \notag \\ &  & \!\!\!\mathbb{P} \left[\! \frac{\rho^t_u \, G_u(0) \, G_H(0) \left(\frac{\nu}{4 \pi}\right)^2 \zeta D_0^{-2} W_S}{\sum_{k=1}^{K} \rho^t_u \, G_u(\theta_k^t) \, G_H(\theta_k^r) \left(\frac{\nu}{4 \pi}\right)^2 \! \zeta D_k^{-2} W_I \!+\!  \sigma^2} \!>\! \gamma_H \! \right], \IEEEeqnarraynumspace
\end{IEEEeqnarray}
where $\sigma^2$ is noise power, $u$ in the subscript of $P_u^{C,Y}$ represents uplink transmission. The parameter $Y$ is defined as $Y\in\{\rm{cll},$\rm{cf}$\}$, where $\rm{cll}$ and $\rm{cf}$ stand for cellular and cell-free networks, respectively.

\section{Uplink Analysis}\label{UA}
\subsection{Cellular Networks}
In this section, we analyze the uplink coverage probability of the typical HAPS in cellular networks, as shown in the right of Fig.~\ref{fig:Figure1_1}. First, considering that users are associated with the HAPS that provides the strongest average received power, the distribution of users served by the typical HAPS is inhomogeneous. Users far from typical HAPS are more likely to be associated with other HAPS. Therefore, the following Lemma gives the probability that users at different locations are associated with typical HAPS.

\begin{lemma}\label{association}
In cellular networks, for the given typical HAPS with coordinate $(R_H,0,0)$, the probability that a user with coordinate $(R_u,\phi,\theta)$ is associated with this HAPS is given as,
\begin{IEEEeqnarray}{RCL} \label{assouplink}
    &&\!\!\!\! P_u^{A,{\rm{cll}}} \left( \theta \right) = \notag \\
    &&\!\!\!\! \exp \left(\! -\! \left(\lambda_H \!-\! \frac{1}{2\pi R_H (R_H\!-\!R_u)}\right) 2 \pi R_H^2 \left(1-\cos\theta\right) \right), \IEEEeqnarraynumspace
\end{IEEEeqnarray}
where $\theta \leq \arccos \frac{R_u}{R_H}$ is satisfied to ensure that the user is in the typical HAPS' LoS region.
\begin{proof}
See Appendix~\ref{app:association}.
\end{proof}
\end{lemma}

The last Lemma studies the distribution of associated users, whereas the following three Lemmas are about interfering users. We first introduce the operator $\mathcal{T}$ to describe the beam deflection angle.

\begin{lemma}\label{angle}
In a triangle with vertices $(R_1,\phi_1,\theta_1)$, $(R_2,\phi_2,\theta_2)$, and $(R_3,\phi_3,\theta_3)$, the interior angle corresponding to vertex $(R_2,\phi_2,\theta_2)$ is, 
\begin{IEEEeqnarray}{RCL}
   &&\!\!\!\!\mathcal{T}\left( R_1,\phi_1,\theta_1;R_2,\phi_2,\theta_2;R_3,\phi_3,\theta_3 \right)    
    = \arccos  \notag \\ 
    &&\!\!\!\!\! \Bigg(\frac{d^2\left( R_1,\phi_1,\theta_1;R_2,\phi_2,\theta_2 \right ) \!+ \!d^2\left( \!\! \!R_2,\phi_2,\theta_2;R_3,\phi_3,\theta_3 \right )} {2d\left( R_1,\phi_1,\theta_1;R_2,\phi_2,\theta_2 \right )d\left( R_2,\phi_2,\theta_2;R_3,\phi_3,\theta_3 \right )} \notag \\
    && \!\!\!\! \!-\! \frac{ d^2\left( R_1,\phi_1,\theta_1;R_3,\phi_3,\theta_3 \right )} {2d\left( R_1,\phi_1,\theta_1;R_2,\phi_2,\theta_2 \right )d\left( R_2,\phi_2,\theta_2;R_3,\phi_3,\theta_3 \right )}\Bigg), \IEEEeqnarraynumspace
\end{IEEEeqnarray}
where $d \left( R_1,\phi_1,\theta_1; R_2,\phi_2,\theta_2 \right )$ represents the Euclidean distance between vertex $(R_1,\phi_1,\theta_1)$ and vertex $(R_2,\phi_2,\theta_2)$, which is given by
\begin{IEEEeqnarray}{RCL}\label{Euclidean}
    d ( R_1, && \phi_1, \theta_1;R_2,\phi_2,\theta_2  )  = 
     \big( R_1^2 + R_2^2 - 2  R_1  R_2   \notag  \\ & &\times (\sin\theta_1\sin\theta_2\cos(\phi_1-\phi_2)+\cos\theta_1 \cos\theta_2) \big)^{\frac{1}{2}}. \IEEEeqnarraynumspace
\end{IEEEeqnarray}
\begin{proof}
We first convert the coordinates from the spherical coordinates to cartesian coordinates and calculate the Euclidean distance. Then, the interior angle is obtained through the law of cosines. 
\end{proof}
\end{lemma}

In the scenario of uplink interference analysis, we consider that the coordinate of typical HAPS is $(R_H,0,0)$, the coordinate of the interfering user is $(R_u,\phi_2,\theta_2)$, and the coordinate of HAPS which is associated with the interfering user is $(R_H,\phi_3,\theta_3)$. Then, the operator $\mathcal{T}\left( R_H,0,0;R_u,\phi_2,\theta_2;R_H,\phi_3,\theta_3 \right)$ outputs the beam deviation angle between the typical HAPS and the alignment beam direction of the interfering user. With this operator, we can calculate the average antenna gain from the user at a given position $(R_u,\phi,\theta)$ to the typical HAPS.

\begin{lemma}\label{gain1}
In cellular networks, given that the coordinate of the typical HAPS is $(R_H,0,0)$, the average antenna gain of uplink transmission from the user with coordinate $(R_u,\phi,\theta)$ can be derived as,
\begin{IEEEeqnarray}{RCL}\label{gainccl}
     \overline{G}_u^{\rm{cll}} && \left( \theta \right) = P_u^{A,{\rm{cll}}} (\theta) G_u(0) \notag \\ 
    && + \int_0^{2\pi} \int_0^\theta \left(\lambda_H - \frac{1}{2\pi R_H (R_H-R_u)}\right) R_H^2  P_u^{A,{\rm{cll}}}(\psi) \notag  \IEEEeqnarraynumspace \\
    &&\times \sin\psi G_u\left( \mathcal{T}\left ( R_H,0,\theta;R_u,0,0;R_H,\varphi,\psi \right) \right) \mathrm{d}\psi \mathrm{d}\varphi,
\end{IEEEeqnarray}
where the association probability $P_u^{A,{\rm{cll}}} (\theta)$ is given in Lemma~\ref{association}.
\begin{proof}
See Appendix~\ref{app:gain1}.
\end{proof}
\end{lemma}

According to the average antenna gain of users at given positions given by Lemma~\ref{gain1}, we can obtain the distribution of interference signal power of users at given positions. The total interference power can be obtained by traversing the entire LoS range of the typical HAPS. As an important intermediate result in the analytical expression of coverage probability, the Laplace transform of interference is further derived in the following lemma. Note that the results in Lemma~\ref{association} and Lemma~\ref{gain1} are independent of the user's azimuth angle, the Laplace transform of the interference power is also not a function of the azimuth angle.

\begin{lemma}\label{LT1}
In cellular networks, given that the typical HAPS at $(R_H,0,0)$ is serving a user with coordinate $(R_u,\phi,\theta)$, the Laplace transform of uplink interfering power caused by other users is given by,
\begin{IEEEeqnarray}{RCL}
     \mathcal{L}_u^{\rm{cll}} (s,\theta && )  =  \int_0^{2\pi} \int_0^{\Theta} m^m \cdot \bigg( m+s\rho_u^t \overline{G}_u^{\rm{cll}}(\psi)  \notag  \\ 
    && \times G_H \big(\mathcal{T} \left( R_u,0,\theta;R_H,0,0;R_u,\varphi,\psi \right) \big) \notag  \\
    && \times \left(\frac{\nu}{4\pi}\right)^2 \zeta \left(d(R_H,0,0;R_u,\varphi,\psi) \right)^{-2} \bigg)^{-m} \notag  \\
    && \times \left( \lambda_u - \frac{{R_H}}{2\pi R_u^2 ({R_H}-{R_u})}\right) R_u^2 \sin\psi \mathrm{d}\varphi \mathrm{d}\psi, \IEEEeqnarraynumspace
\end{IEEEeqnarray}
where $\overline{G}_u^{\rm{cll}}(\psi)$ is defined in Lemma~\ref{gain1}, $\mathcal{T}\left( R_u,0,\theta;R_H,0,0;R_u,\varphi,\psi \right)$ and $d(R_H,0,0;R_u,\varphi,\psi)$ are defined in Lemma~\ref{angle}.
\begin{proof}
See Appendix~\ref{app:LT1}.
\end{proof}
\end{lemma}

Based on the above lemmas, the coverage probability is given by the following theorem.
\begin{theorem}\label{cov1}
For a given typical HAPS, the uplink coverage probability in cellular networks is given by,
\begin{IEEEeqnarray}{RCL}\label{cov11}
     P_u^{C,{\rm{cll}}} && (\gamma_H)  = \int_0^{\Theta} \Bigg( 1 - \left( \frac{2b_0 n}{2 b_0 n + \Omega} \right)^n  \notag  \\
     \times &&\sum_{z=0}^{\infty} \frac{(n)_z}{z!} \!\left(\! \frac{\Omega}{2 b_0 n + \Omega} \right)^{\!z}  \! \bigg( 1 \! - \!\sum_{k=1}^{z+1} \binom{z+1}{k} (-1)^{k+1} \notag  \IEEEeqnarraynumspace \\
     \times && \exp\left(-k\beta_u(\theta,\gamma_H) \sigma^2 \right) \mathcal{L}_u^{\rm{cll}}(k\beta_u(\theta,\gamma_H),\theta) \bigg) \Bigg) \notag \\
     \times && \lambda_H  2 \pi R_H^2 \sin\theta \exp \left( -{\lambda}_H 2 \pi R_H^2 \left(1-\cos\theta\right) \right) \mathrm{d}\theta,
\end{IEEEeqnarray}
where $\mathcal{L}_u^{\rm{cll}} (k\beta_u(\theta,\gamma_H), \theta)$ is the Laplace transform of the interference power defined in Lemma~\ref{LT1}, $P_u^{A,{\rm{cll}}}(\theta)$ is the association probability defined in Lemma~\ref{association}, and $\beta_u(\theta,\gamma_H)$ is defined as,
\begin{IEEEeqnarray}{RCL}\label{betau}
    \beta_u(\theta,\gamma_H)  =& & \big((z+1)!\big)^{\frac{-1}{z+1}} \notag \\
    && \times \frac{\gamma_H \times \left(R_H^2 + R_u^2 -2 R_H R_u \cos\theta\right) }{2b_0 \rho_u^t G_u(0) G_H(0) \left(\frac{\nu}{4\pi}\right)^2 \zeta }.
\IEEEeqnarraynumspace
\end{IEEEeqnarray}
\begin{proof}
See Appendix~\ref{app:cov1}.
\end{proof}
\end{theorem}

\subsection{Cell-Free Networks}
In this section, we analyze the coverage probability of the typical HAPS in cell-free networks, as shown in the left of Fig.~\ref{fig:Figure1_1}. In general, the analytical framework of cell-free networks is similar to that of cellular networks. Compared with cellular networks, cell-free networks have a simpler expression with regard to the association relationship between the typical HAPS and users.

\begin{lemma}\label{gain2}
In cell-free networks, given that the coordinate of the typical HAPS is $(R_H,0,0)$, the average antenna gain of uplink transmission from the user with coordinate $(R_u,\phi,\theta)$ can be derived as,
\begin{IEEEeqnarray}{RCL} \label{gaincf}
    \overline{G}_u^{\rm{cf}} & & \left( \theta \right)  = \frac{G_u(0)}{\lambda_H 2 \pi R_H \left( R_H - R_u \right)} \notag   \\ 
    && + \int_0^{2\pi} \int_0^{\Theta} \left( \lambda_H - \frac{1}{2 \pi R_H \left( R_H - R_u \right)}  \right)  R_H^2 \sin\psi \notag \IEEEeqnarraynumspace \\
    && \times   \frac{G_u\left( \mathcal{T}\left (R_H,0,\theta; R_u,0,0; R_H,\varphi,\psi \right) \right)}{\lambda_H 2 \pi R_H \left( R_H - R_u \right)}  \mathrm{d}\psi \mathrm{d}\varphi,
\end{IEEEeqnarray}
where $\mathcal{T}\left (R_H,0,\theta; R_u,0,0; R_H,\varphi,\psi \right)$ is defined in Lemma~\ref{angle}.
\begin{proof}
See Appendix~\ref{app:gain2}.
\end{proof}
\end{lemma}

\begin{lemma}\label{LT2}
In cell-free networks, given that the typical HAPS at $(R_H,0,0)$ is serving a user with coordinate $(R_u,\phi,\theta)$, the Laplace transform of uplink interfering power caused by users is given by,
\begin{IEEEeqnarray}{RCL}
    \mathcal{L}_u^{\rm{cf}} ( s,&& \theta)  = \int_0^{2\pi} \int_0^{\Theta} m^m \cdot \bigg( m+s\rho_u^t \overline{G}_u^{\rm{cf}}(\psi) \notag \\ 
    && \times G_H\big(\mathcal{T}\left( R_u,0,\theta;R_H,0,0;R_u,\varphi,\psi \right) \big) \notag  \\
    && \times \left(\frac{\nu}{4\pi}\right)^2 \zeta \left(d(R_H,0,0;R_u,\varphi,\psi) \right)^{-2} \bigg)^{-m} \notag  \\
    && \times \left( \lambda_u - \frac{{R_H}}{2\pi R_u^2 ({R_H}-{R_u})}\right) R_u^2 \sin\psi \mathrm{d}\varphi \mathrm{d}\psi, \IEEEeqnarraynumspace
\end{IEEEeqnarray}
where $\overline{G}_u^{\rm{cf}}(\psi)$ is defined in Lemma~\ref{gain2}, $\mathcal{T}( R_u,0,\theta;R_H,0,0;$ $ R_u,\varphi,\psi )$ and $d(R_H,0,0;R_u,\varphi,\psi)$ are defined in Lemma~\ref{angle}.
\begin{proof}
The proof of Lemma~\ref{LT2} is similar to that of Lemma~\ref{LT1}, therefore omitted here.
\end{proof}
\end{lemma}

\begin{theorem}\label{cov2}
For a given typical HAPS, the uplink coverage probability in cell-free networks is given by,
\begin{IEEEeqnarray}{RCL} \label{cov22}
     &&P_u^{C,{\rm{cf}}} (\gamma_H)  = \int_0^{\Theta} \frac{R_H \sin\theta}{R_H-R_u} \Bigg( 1 - \left( \frac{2b_0 n}{2 b_0 n + \Omega} \right)^n \notag  \\ 
    &&  \times \sum_{z=0}^{\infty} \frac{(n)_z}{z!} \left( \frac{\Omega}{2 b_0 n + \Omega} \right)^z  \bigg( 1 \!-\! \sum_{k=1}^{z+1} \binom{z+1}{k} (-1)^{k+1} \notag \IEEEeqnarraynumspace \\
    && \times  \exp\left(-k\beta_u(\theta,\gamma_H) \sigma^2 \right) \mathcal{L}_u^{\rm{cf}}\big(k\beta_u(\theta,\gamma_H),\theta\big) \bigg) \Bigg)  \mathrm{d}\theta,
\end{IEEEeqnarray}
where $\mathcal{L}_u^{\rm{cf}} (k\beta_u(\theta,\gamma_H), \theta)$ is the Laplace transform of the interference power defined in Lemma~\ref{LT2}, and $\beta_u(\theta,\gamma_H)$ is defined in (\ref{betau}).
\begin{proof}
In cell-free networks, the probability of providing service to users in any area is the same because it is within the LoS region of the typical HAPS. Therefore, the PDF of the contact angle distribution, which is the distribution of the central angle between the typical HAPS and serving user, is given as,
\begin{IEEEeqnarray}{RCL}\label{contacf}
    f_{\theta_0} (\theta) & =&  \frac{\mathrm{d}}{\mathrm{d}\theta} F_{\theta_0} (\theta) \notag\\ 
    &=&  \frac{\mathrm{d}}{\mathrm{d}\theta} \frac{2\pi R_u^2 (1-\cos\theta)}{2\pi R_u^2 (1-\cos\Theta)} = \frac{\sin\theta}{1-\cos\Theta},
\end{IEEEeqnarray}
where $\Theta=\arccos({R_u}/{R_H})$ is the upper bound of the central angle. After replacing the results of (\ref{contacf}) into (\ref{contactangle}), the remaining proof of Theorem~\ref{cov2} is similar to that of Theorem~\ref{cov1}, therefore omitted here.
\end{proof}
\end{theorem}

\vspace{-0.2cm}
\section{Downlink Analysis} \label{DA}
This section analyzes the downlink coverage probability of the typical user. The analysis of downlink coverage probability is similar to that of uplink coverage probability. Due to space limitations, the analytic results are extended from the uplink scenario and most of the proofs are omitted.

\vspace{-0.2cm}
\subsection{Cellular Networks}
\begin{lemma}\label{gain3}
In cellular networks, given that the coordinate of the typical user is $(R_u,0,0)$, the approximate average antenna gain of downlink transmission from the interfering HAPS with coordinate $(R_H,\phi,\theta)$ can be estimated as,
\begin{IEEEeqnarray}{RCL}\label{gainccl3}
     \overline{G}_H^{\rm{cll}} \left( \theta \right) &&= \int_0^{2\pi} \int_0^{\Theta}      \frac{\sin\psi P_u^A(\psi) }{2\pi \int_0^{\Theta} \sin\psi P_u^A(\psi) \mathrm{d}\psi} \notag\\ 
     && \times  G_H\left( \mathcal{T}\left ( R_u,0,\theta;R_H,0,0;R_u,\varphi,\psi \right) \right)  \mathrm{d}\psi \mathrm{d}\varphi,
\IEEEeqnarraynumspace
\end{IEEEeqnarray}
where the association probability $P_u^{A,{\rm{cll}}} (\theta)$ is defined in Lemma~\ref{association}.
\begin{proof}
See Appendix~\ref{app:gain3}.
\end{proof}
\end{lemma}

\begin{lemma}\label{LT3}
In cellular networks, given that the typical user at $(R_u,0,0)$ is served by a HAPS with coordinate $(R_u,\phi,\theta)$, the Laplace transform of downlink interfering power from other interfering HAPS is given by,
\begin{IEEEeqnarray}{RCL}
     \mathcal{L}_d^{\rm{cll}}  && (s,\theta) = \int_0^{2\pi} \int_\theta^{\Theta} m^m \cdot \bigg( m+s\rho_H^t \overline{G}_H^{\rm{cll}}(\psi) \notag\\
     && \times G_u\big(\mathcal{T}\left( R_H,0,\theta;R_u,0,0;R_H,\varphi,\psi \right) \big)  \notag \\
    && \times \left(\frac{\nu}{4\pi}\right)^2 \zeta \big(d(R_u,0,0;R_H,\varphi,\psi) \big)^{-2} \bigg)^{-m} 
    \notag \\
    && \times \left( \lambda_H - \frac{1}{2\pi R_H (R_H-R_u)}\right) R_H^2 \sin\psi \mathrm{d}\varphi \mathrm{d}\psi, \IEEEeqnarraynumspace
\end{IEEEeqnarray}
where $\overline{G}_H^{\rm{cll}}(\psi)$ is defined in Lemma~\ref{gain3}, $\mathcal{T}( R_H,0,\theta;R_u,0,0;$ $R_H,\varphi,\psi )$ and $d(R_u,0,0;R_H,\varphi,\psi)$ are defined in Lemma~\ref{angle}.
\begin{proof}
In cellular networks, interfering HAPS are further away from the typical user than the associated HAPS. Therefore, the integral interval of $\psi$ needs to be changed accordingly. The remaining proof of Lemma~\ref{LT3} is similar to that of Lemma~\ref{LT1}, therefore, it is omitted here.
\end{proof}
\end{lemma}

\begin{theorem}\label{cov3}
For a given typical user, the downlink coverage probability in cellular networks is given by,
\begin{IEEEeqnarray}{RCL}\label{cov33}
     && P_d^{C,{\rm{cll}}} (\gamma_u)  = \int_0^{\Theta} \Bigg( 1 - \left( \frac{2b_0 n}{2 b_0 n + \Omega} \right)^n \sum_{z=0}^{\infty} \frac{(n)_z}{z!} \notag\\ 
    && \times \left( \frac{\Omega}{2 b_0 n + \Omega} \right)^z  \bigg( 1 - \sum_{k=1}^{z+1} \binom{z+1}{k} (-1)^{k+1} \notag \\ 
    & &\times \exp\big(-k\beta_H(\theta,\gamma_u) \sigma^2 \big) \mathcal{L}_d^{\rm{cll}}\big(k\beta_H(\theta,\gamma_u),\theta\big) \bigg) \Bigg) \notag \\
    && \times  \lambda_H  2 \pi R_H^2 \sin\theta \exp \left( -{\lambda}_H 2 \pi R_H^2 \left(1-\cos\theta\right) \right) \mathrm{d}\theta,
\IEEEeqnarraynumspace
\end{IEEEeqnarray}
where $\mathcal{L}_d^{\rm{cll}} (k\beta_H
(\theta,\gamma_u), \theta)$ is the Laplace transform of the interference power defined in Lemma~\ref{LT3}, $P_u^{A,{\rm{cll}}}(\theta)$ is the association probability defined in Lemma~\ref{association}, and $\beta_H(\theta,\gamma_u)$ is defined as,
\begin{IEEEeqnarray}{RCL}\label{betaH}
     \beta_H(\theta,\gamma_u)  = && \big((z+1)!\big)^{\frac{-1}{z+1}} \notag \\ 
     && \times \frac{\gamma_u  \left(R_H^2 + R_u^2 -2 R_H R_u \cos\theta\right) }{2b_0 \rho_H^t G_u(0) G_H(0) \left(\frac{\nu}{4\pi}\right)^2 \zeta }.
\end{IEEEeqnarray}
\begin{proof}
The proof of Theorem~\ref{cov3} is similar to that of Theorem~\ref{cov1}, therefore omitted here.
\end{proof}
\end{theorem}

\begin{figure}[t]
	\centering
    \includegraphics[width=0.95\linewidth]{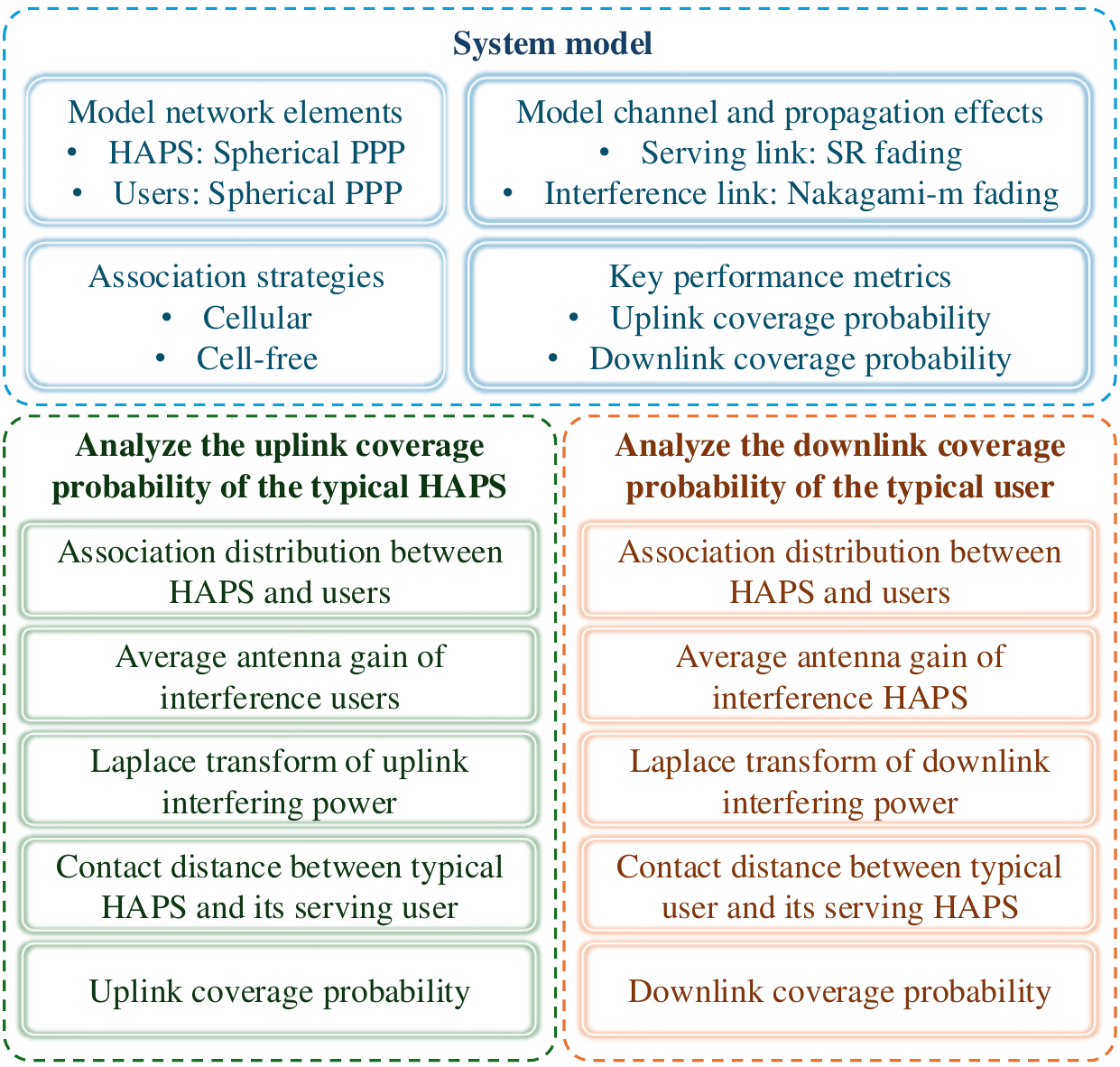}
	\caption{Block diagram to describe the overall evaluation effectiveness of SG.}
	\label{fig:Figureblock}
	\vspace{-0.5cm}
\end{figure}

\vspace{-0.4cm}
\subsection{Cell-Free Networks}
\begin{lemma}\label{gain4}
In cell-free networks, given that the coordinate of the typical user is $(R_u,0,0)$, the average antenna gain of the downlink transmission from the interfering HAPS with coordinate $(R_H,\phi,\theta)$ can be derived as,
\begin{IEEEeqnarray}{RCL}\label{gaincf4}
   \overline{G}_H^{\rm{cf}}  &&\left( \theta \right)  =  \int_0^{2\pi} \!\!  \int_0^{\Theta} \left( \lambda_u \!-\! \frac{{R_H}}{2\pi R_u^2 ({R_H}-{R_u})}\right)  R_u^2 \sin\psi  \notag \\
    && \times \frac{R_H G_u\left( \mathcal{T}\left (R_u,0,\theta; R_H,0,0; R_u,\varphi,\psi \right) \right)}{\lambda_u 2 \pi R_u^2 \left( R_H - R_u \right)} \mathrm{d}\psi \mathrm{d}\varphi.
\IEEEeqnarraynumspace
\end{IEEEeqnarray}
\begin{proof}
The proof of Lemma~\ref{gain4} is similar to that of Lemma~\ref{gain2}, therefore omitted here. 
\end{proof}
\end{lemma}

\begin{figure*}[t!]
\begin{minipage}[t]{0.328\linewidth}
\centering
\includegraphics[width=0.99\linewidth]{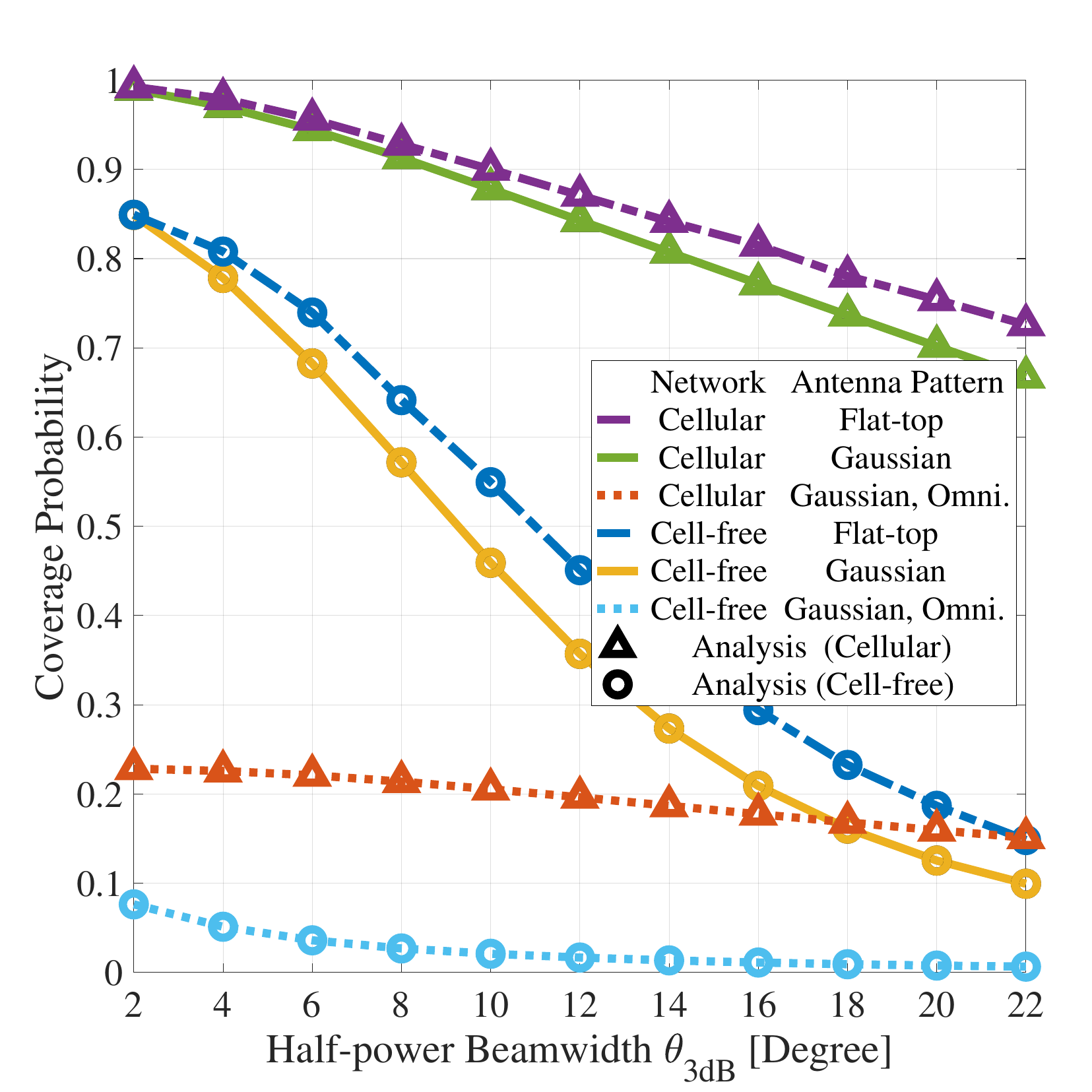}
\caption{Influence of different antenna patterns on uplink coverage probability.}
\label{fig1}
\end{minipage}
\hfill
\begin{minipage}[t]{0.328\linewidth}
\centering
\includegraphics[width=0.99\linewidth]{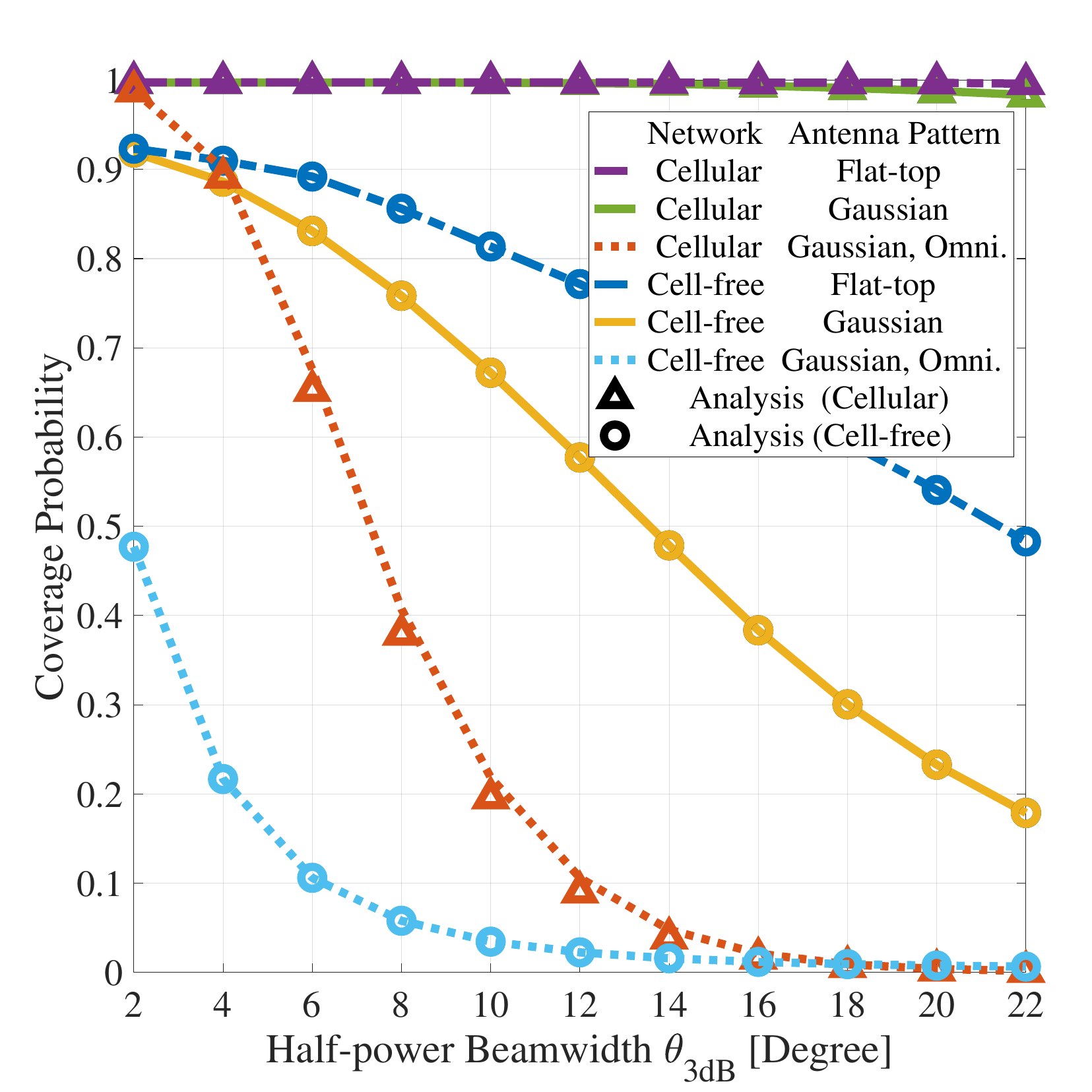}
\caption{Influence of different antenna patterns on downlink coverage probability.} 
\label{fig2}
\end{minipage}
\hfill
\begin{minipage}[t]{0.328\linewidth}
\centering
\includegraphics[width=0.99\linewidth]{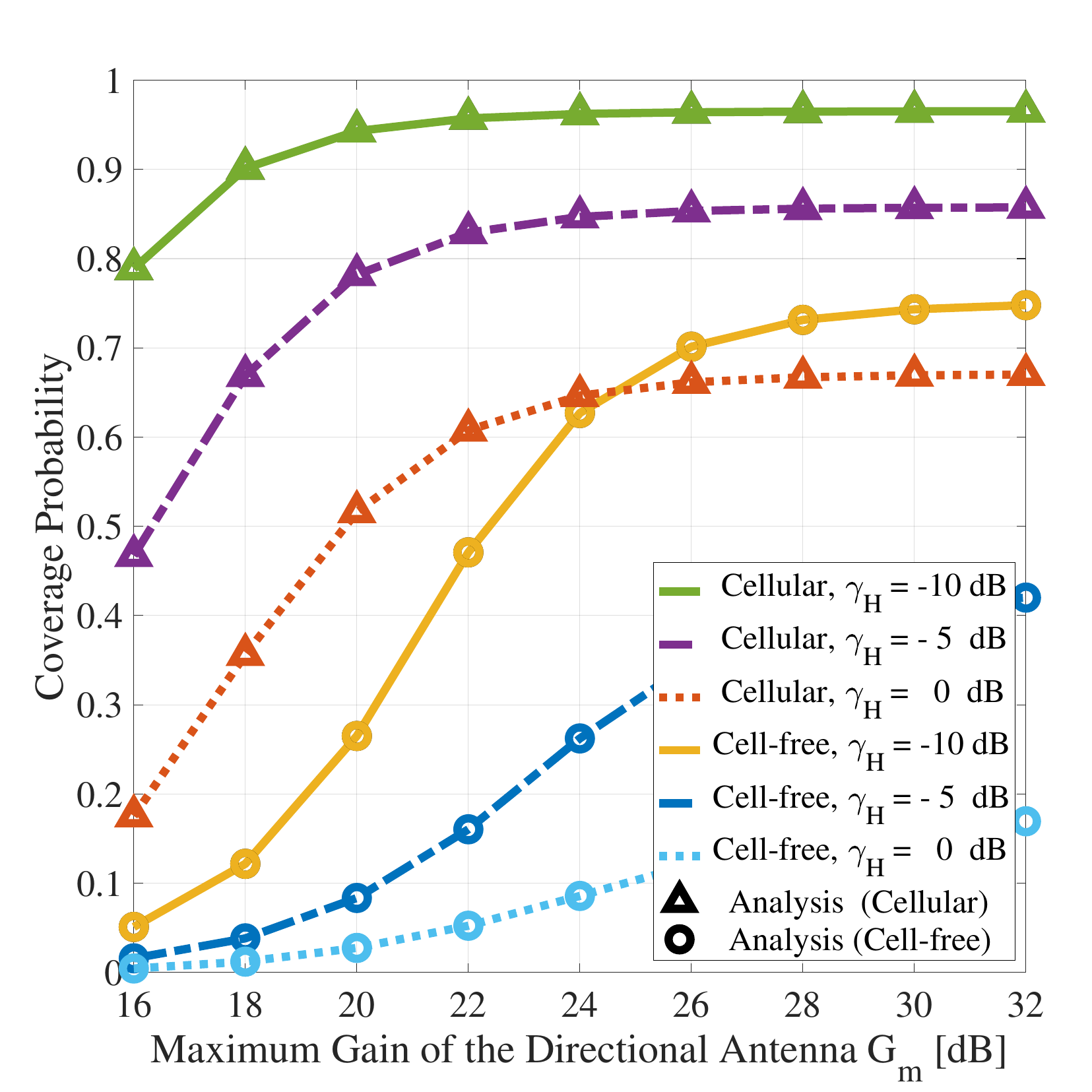}
\caption{Influence of maximum antenna gain on uplink coverage probability.}
\label{fig3}
\end{minipage}
\vspace{-0.4cm}
\end{figure*}

\vspace{-0.2cm}
\begin{lemma}\label{LT4}
In cell-free networks, given that the typical user at $(R_u,0,0)$ is served by a HAPS with coordinate $(R_H,\phi,\theta)$, the Laplace transform of downlink interfering power is given by,
\vspace{-0.3cm}
\begin{IEEEeqnarray}{RCL}
     \mathcal{L}_H^{\rm{cf}} &&(s,\theta) = \int_0^{2\pi} \int_0^{\Theta} m^m \cdot \bigg( m+s\rho_H^t \overline{G}_H^{\rm{cf}}(\psi)  \notag \\
    && \times G_u\big(\mathcal{T}\left( R_H,0,\theta;R_u,0,0;R_H,\varphi,\psi \right) \big) \notag \\
    && \times \left(\frac{\nu}{4\pi}\right)^2 \zeta \big(d(R_u,0,0;R_H,\varphi,\psi) \big)^{-2} \bigg)^{-m} \notag \\
    && \times\left( \lambda_H - \frac{1}{2\pi R_H (R_H-R_u)}\right) R_H^2 \sin\psi \mathrm{d}\varphi \mathrm{d}\psi,
\IEEEeqnarraynumspace
\end{IEEEeqnarray}
where $\overline{G}_H^{\rm{cf}}(\psi)$ is defined in Lemma~\ref{gain3}, $\mathcal{T}( R_H,0,\theta;R_u,0,0;$ $R_H,\varphi,\psi )$ and $d(R_u,0,0;R_H,\varphi,\psi)$ are defined in Lemma~\ref{angle}.
\begin{proof}
The proof of Lemma~\ref{LT4} is similar to that of Lemma~\ref{LT1}, therefore omitted here.
\end{proof}
\end{lemma}

\vspace{-0.2cm}
\begin{theorem}\label{cov4}
For a given typical user, the downlink coverage probability in cell-free networks is given by,
\begin{IEEEeqnarray}{RCL}\label{cov44}
     && P_d^{C,{\rm{cf}}} (\gamma_u)  = \int_0^{\Theta} \frac{R_H \sin\theta}{R_H-R_u} \Bigg( 1 - \left( \frac{2b_0 n}{2 b_0 n + \Omega} \right)^n  \notag \\
    && \times  \sum_{z=0}^{\infty} \frac{(n)_z}{z!} \left( \frac{\Omega}{2 b_0 n + \Omega} \right)^z  \bigg( 1 \!- \!\sum_{k=1}^{z+1} \binom{z+1}{k} (-1)^{k+1}  \notag  \IEEEeqnarraynumspace \\
    && \times \exp\left(-k\beta_H(\theta,\gamma_u) \sigma^2 \right) \mathcal{L}_H^{\rm{cf}}(k\beta_H(\theta,\gamma_u),\theta) \bigg) \Bigg)  \mathrm{d}\theta,
\end{IEEEeqnarray}
where $\mathcal{L}_H^{\rm{cf}} (k\beta_H(\theta,\gamma_u), \theta)$ is the Laplace transform of the interference power defined in Lemma~\ref{LT4}, and $\beta_H(\theta,\gamma_u)$ is defined in (\ref{betaH}).
\begin{proof}
The remaining proof of Theorem~\ref{cov4} is similar to that of Theorem~\ref{cov2}, therefore omitted here.
\end{proof}
\end{theorem}

It is important to emphasize that the expressions for coverage probability ((\ref{cov11}),  (\ref{cov22}), (\ref{cov33}), and  (\ref{cov44})) cannot be straightforwardly expressed in closed-form. This difficulty arises from the complexity of spherical topology, as the curvature of the sphere complicates geometric relationships. Furthermore, considering interference further adds to the challenge of simplifying these expressions.


\subsection{Overall Evaluation Effectiveness}

As shown in Fig. \ref{fig:Figureblock}, the evaluation process begins with the establishment of a system model, encompassing network elements, association strategies, channel models, and a parameter matrix. Firstly, the primary distinction between the uplink and downlink coverage probability evaluation lies in the different sources of interference. Specifically, each user can associate with only one HAPS, whereas a single HAPS can associate with multiple users. This results in a more complex interference analysis for the uplink compared to the downlink. Secondly, the analytical framework for cellular networks is more intricate than that for cell-free networks. This is because the association distribution between HAPS and users is inhomogeneous in cellular networks, whereas it is homogeneous in cell-free networks.

\section{Numerical Results} \label{NM}
\begin{figure*}[htbp]
\begin{minipage}[t]{0.328\linewidth}
\centering
\includegraphics[width=0.99\linewidth]{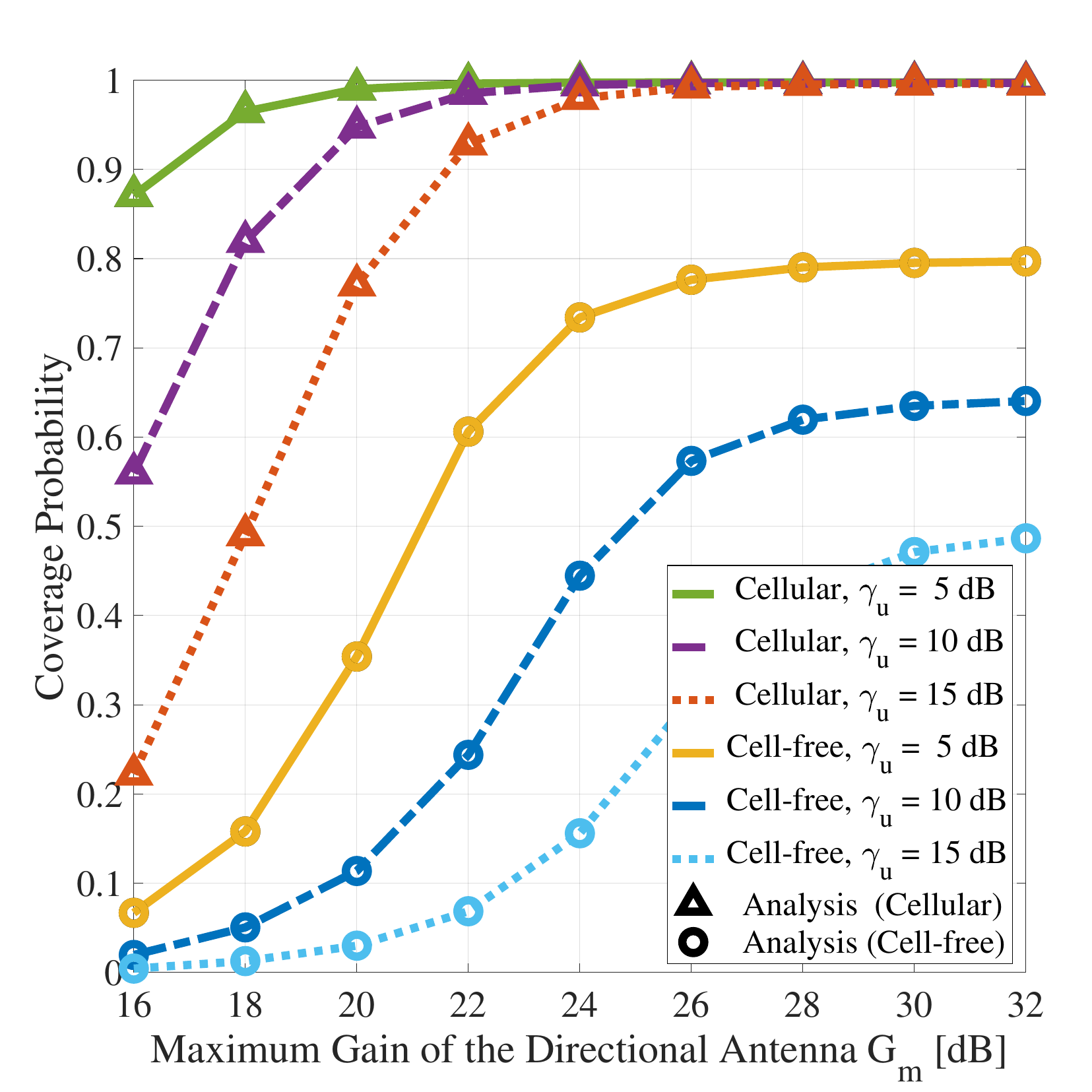}
\caption{Influence of maximum antenna gain on downlink coverage probability.} 
\label{fig4}
\end{minipage}
\hfill
\begin{minipage}[t]{0.328\linewidth}
\centering
\includegraphics[width=0.99\linewidth]{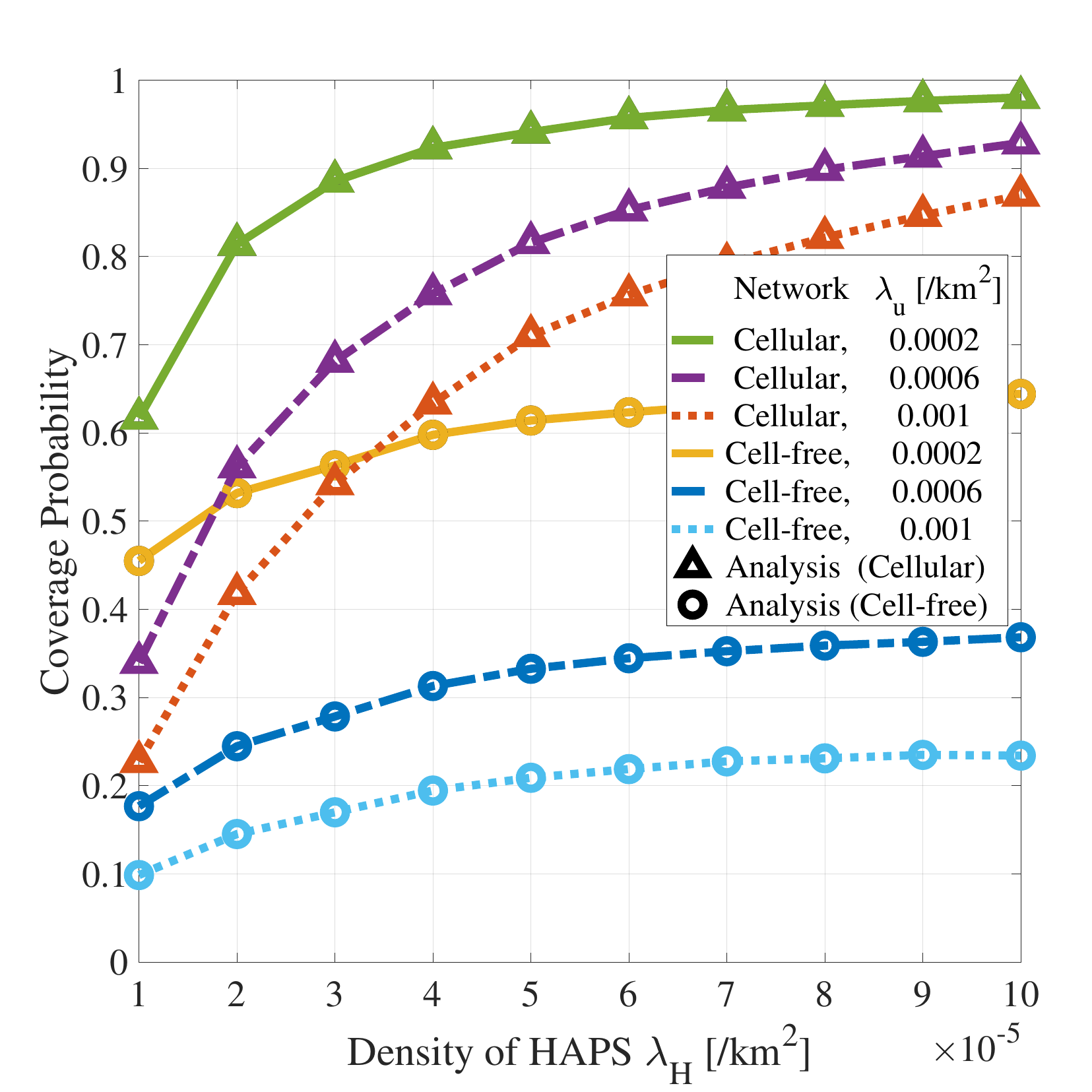}
\caption{Influence of density of HAPS and users on uplink coverage probability.}
\label{fig5}
\end{minipage}
\hfill
\begin{minipage}[t]{0.328\linewidth}
\centering
\includegraphics[width=0.99\linewidth]{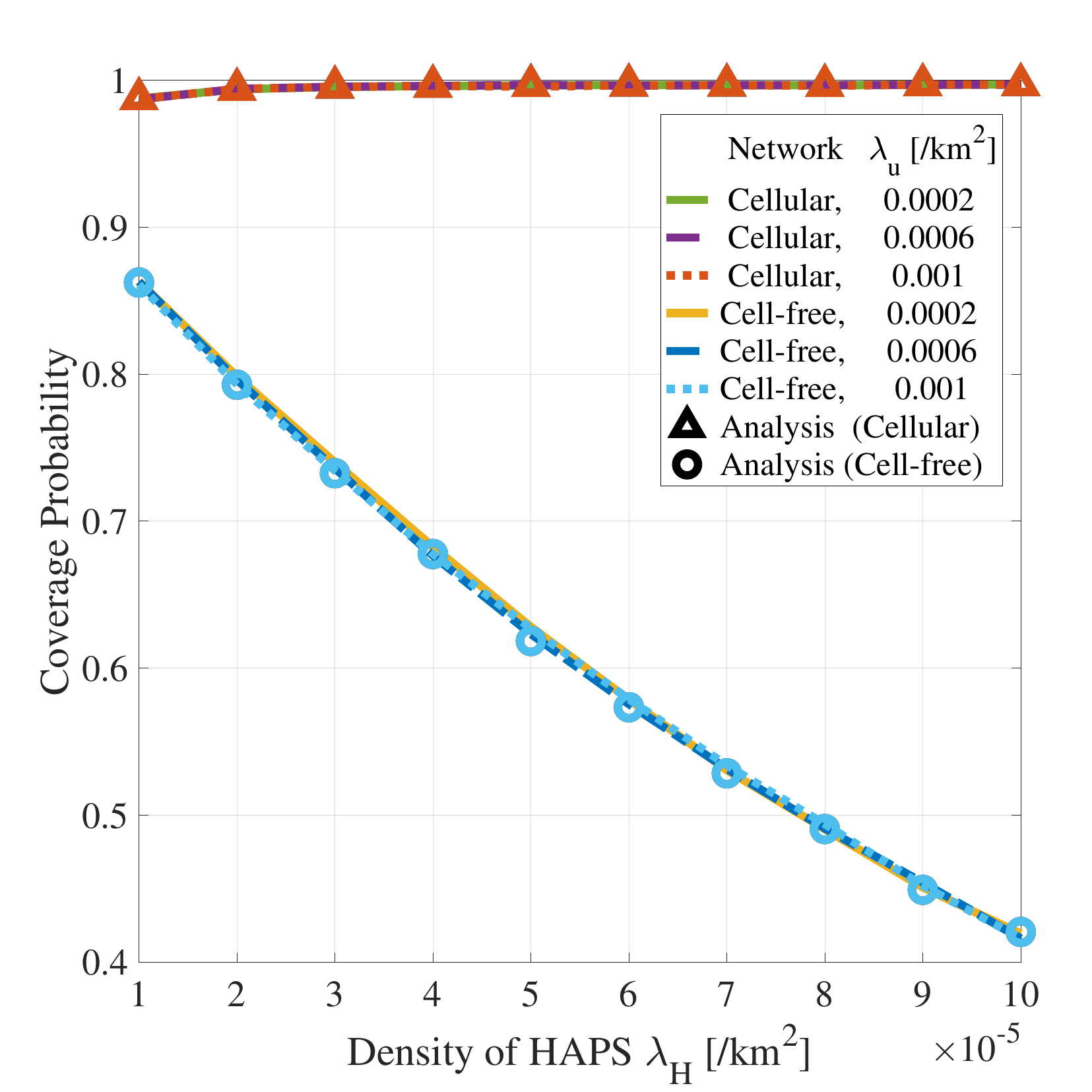}
\caption{Influence of density of HAPS and users on downlink coverage probability.} 
\label{fig6}
\end{minipage}
\vspace{-0.4cm}
\end{figure*}

In this section, we analyze the effects of antenna parameters as well as users and HAPS distributions on the uplink coverage probability (Fig.~\ref{fig1}, Fig.~\ref{fig3} and Fig.~\ref{fig5}) and downlink coverage probability (Fig.~\ref{fig2}, Fig.~\ref{fig4} and Fig.~\ref{fig6}) in cellular and cell-free networks. In the figures, we use triangles and circles to represent the analytical results for cellular and cell-free networks, respectively,  whereas lines represent the results obtained by Monte Carlo simulations. 
In simulations, all the metrics used are averaged over 10,000 iterations. According to the law of large numbers, the mean of the simulation results converges to the theoretical expected value. Moreover, the analytical results and simulation outcomes match perfectly, thereby validating the accuracy of the theorems derived in this paper.
The simulation parameters are summarized in Table \ref{table2}, unless stated otherwise. 

\begin{table}[ht]
\centering
\caption{Parameters and default values\cite{talgat2020stochastic,song2022cooperative}.} 
\label{table2}
\resizebox{\linewidth}{!}{
 \renewcommand{\arraystretch}{1.1}
\begin{tabular}{|c|c|c|}
\hline
Notation       & Parameter                               & Value                 \\ \hline \hline
$R_{u}$, $R_{H}$               & Radius of users, HAPS             & 6371, 6391 [km]               \\ \hline
$\lambda_{u}$, $\lambda_{H}$   & Density of users, HAPS            & 6, 0.6 [$\times10^{-4}$~km$^{-2}$]       \\ \hline
$\rho_{u}^t$, $\rho_{H}^t$     & Transmission power of users, HAPS & 3, 15~[dBW]                    \\ \hline
$G_{\rm m}$   & Maximum gain  & $26$~dBi                \\ \hline
$\gamma_u$, $\gamma_H$ & Coverage threshold of users, HAPS  & $10$, $-5$~[dB]      \\ \hline
$\zeta$        & Average rain attenuation                & $-2$~dB                 \\ \hline
$ \mathcal{SR}(\Omega ,b_0,n)$ & Parameters of SR fading                            & $ \mathcal{SR}(1.29,0.158,19.4)$ \\ \hline
$m$          & Parameter of Nakagami-$m$ fading         & $2$                   \\ \hline
$\nu$          & Wavelength of transmission link        & $0.0150$~m              \\ \hline
$\sigma^2$     & Noise power            & $ 1 \times 10^{-12}$~W \\ \hline
\end{tabular}}
\end{table}

\par
When applying Monte Carlo simulation to calculate uplink coverage probability, it suffices to take typical HAPS as the center, generate users in the range of polar angle less than $\Theta$, and generate HAPS in the range of polar angle less than $2\Theta$. We note that $\Theta=\arccos({R_u}/{R_H})$ is the maximum center angle that can be formed between the user and HAPS within the LoS range. Users with polar angles that are less than $\Theta$ may associate with HAPS with polar angle that is up to $2\Theta$. The same remark above applies for the simulation of downlink coverage probability.


\subsection{Antenna Pattern Impact}



In Fig.~\ref{fig1} and Fig.~\ref{fig2}, the impact of various antenna patterns on uplink and downlink coverage probability is depicted. At the transmitters, directional antennas are employed, while receivers use the same directional antennas as the transmitter, except for two scenarios (red and sky blue dot lines) where omnidirectional antennas are adopted.  

We can observe from Fig.~\ref{fig1} and Fig.~\ref{fig2} that employing directional antennas at the receiver can significantly enhance the coverage probability. This is because directional signal reception can reduce interference by selectively focusing on signals from a specific direction while attenuating signals from other directions. It is straightforward to notice that a smaller half-power beamwidth corresponds to a greater coverage probability. This relationship can be explained by the fact that a smaller half-power beamwidth concentrates the signal energy into a narrower beam, directing more of the transmitted power toward the desired direction and minimizing signal spillage into unwanted areas. 

It also shows in Fig.~\ref{fig1} and Fig.~\ref{fig2} that cellular networks outperforms cell-free networks. This is attributed to the fact that in cell-free networks, where users are more likely to be associated with distance HAPS, the received signal strength tends to be weaker due to higher path loss. 
Specifically, in cell-free networks, each HAPS within a user's LoS range has an equal probability of being associated. In contrast, in cellular networks, users are always associated with the nearest HAPS. 

\vspace{-0.2cm}
\subsection{Parameter Selection}


Fig.~\ref{fig3} and Fig.~\ref{fig4} show the coverage probability under different coverage thresholds and antenna gains, with the transmitter and receiver employing Gaussian antenna patterns. We configure different coverage thresholds for uplink and downlink, since the lower transmit power of users results in weaker signal strength received by HAPS. Additionally, HAPS are equipped with a more powerful demodulation system capable of handling lower SINR signals.

We can see from Fig.~\ref{fig3} and Fig.~\ref{fig4} that with the enhancement of antenna gain, the coverage probability initially increases before converging. This is because as antenna gain increases, signal and interference strengths increase accordingly, while noise power remains constant. 

As observed in Fig.~\ref{fig3} and Fig.~\ref{fig4}, cellular networks outperform cell-free networks under the same coverage threshold. This phenomenon is primarily attributed to the fact that, in the cell-free network architecture, the distances between users and HAPS are generally greater than those in cellular networks, as explained earlier. It also indicates that the contact distance (the distance between a user and his associated HAPS) has a greater impact on coverage probability than interference.




\subsection{User and HAPS Density Impact}



Fig.~\ref{fig5} and Fig.~\ref{fig6} depict the influence of user and HAPS density on the coverage probability, with transmitter and receiver employing Gaussian antenna patterns

As we can see from Fig.~\ref{fig5}, enhancing the HAPS density while keeping the user density constant can improve the coverage probability, with a more pronounced enhancement effect observed in cellular networks compared to cell-free networks. The reason for this phenomenon is that with higher HAPS density, the contact distance becomes shorter in cellular networks, while that in cell-free networks does not decrease, due to random associations. Furthermore, reducing user density also improves coverage probability. This is because a lower number of users associated with each HAPS leads to fewer sources of co-frequency interference, consequently, improving the performance. 

Fig.~\ref{fig6} also reflects that cellular networks keep excellent coverage performances. This indicates that users aligned with interfering HAPS' beams have a large distance from the typical user, so the probability of the typical user falling into the interfering HAPS' main lobe is low. On the contrary, the randomness of beam alignment in cell-free networks results in a linear negative correlation between the density of HAPS and the coverage probability. This correlation stems from the limitation that augmenting HAPS density does not effectively reduce the average contact distance between HAPS and users; instead, it tends to exacerbate interference within the network.

\begin{figure*}[t!]
\begin{minipage}[t]{0.328\linewidth}
    \centering
    \includegraphics[width=0.99\linewidth]{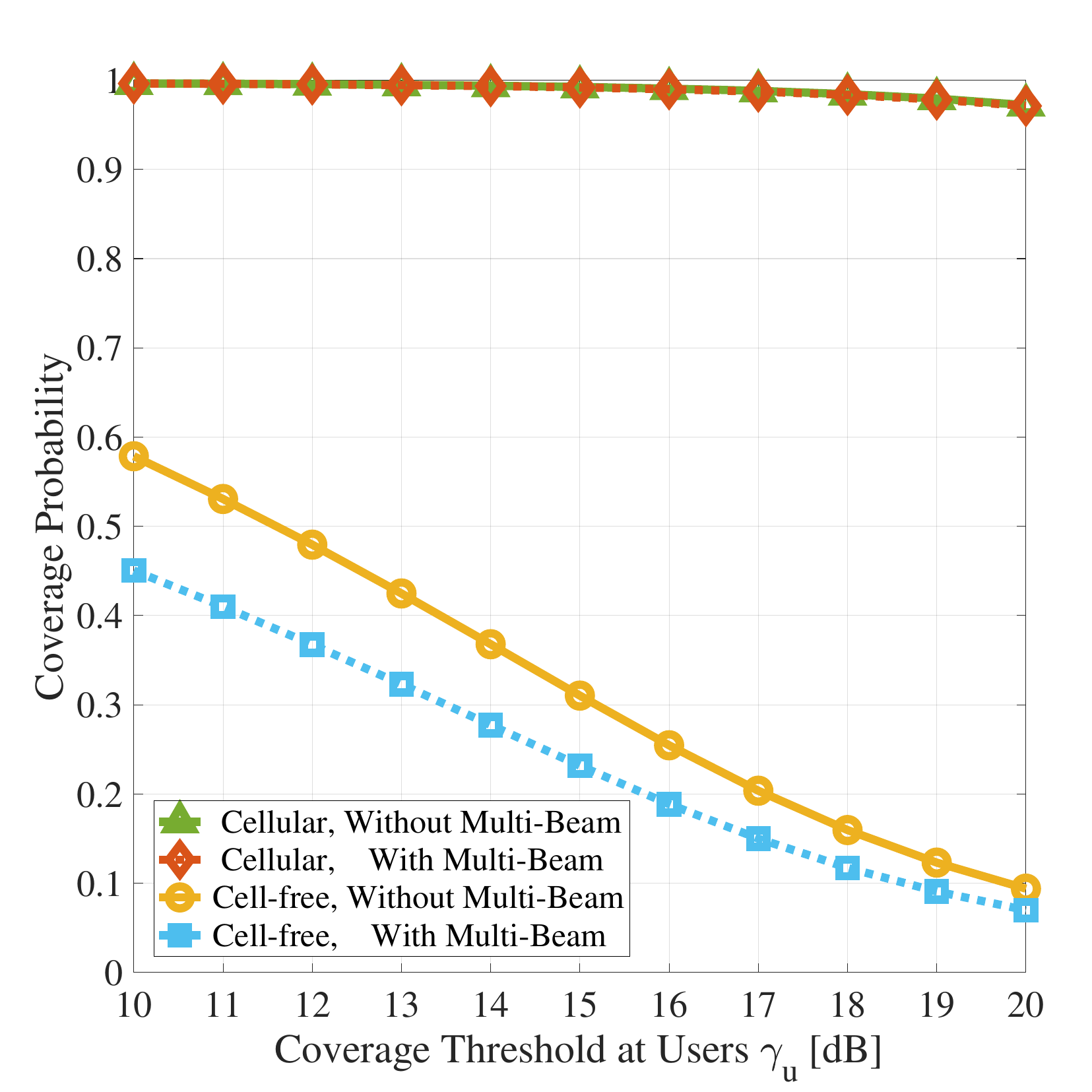}
    \caption{Influence of multi-beam on downlink coverage probability.} 
   \label{fig7}
\end{minipage}
\hfill
\begin{minipage}[t]{0.328\linewidth}
 \centering
\includegraphics[width=0.99\linewidth]{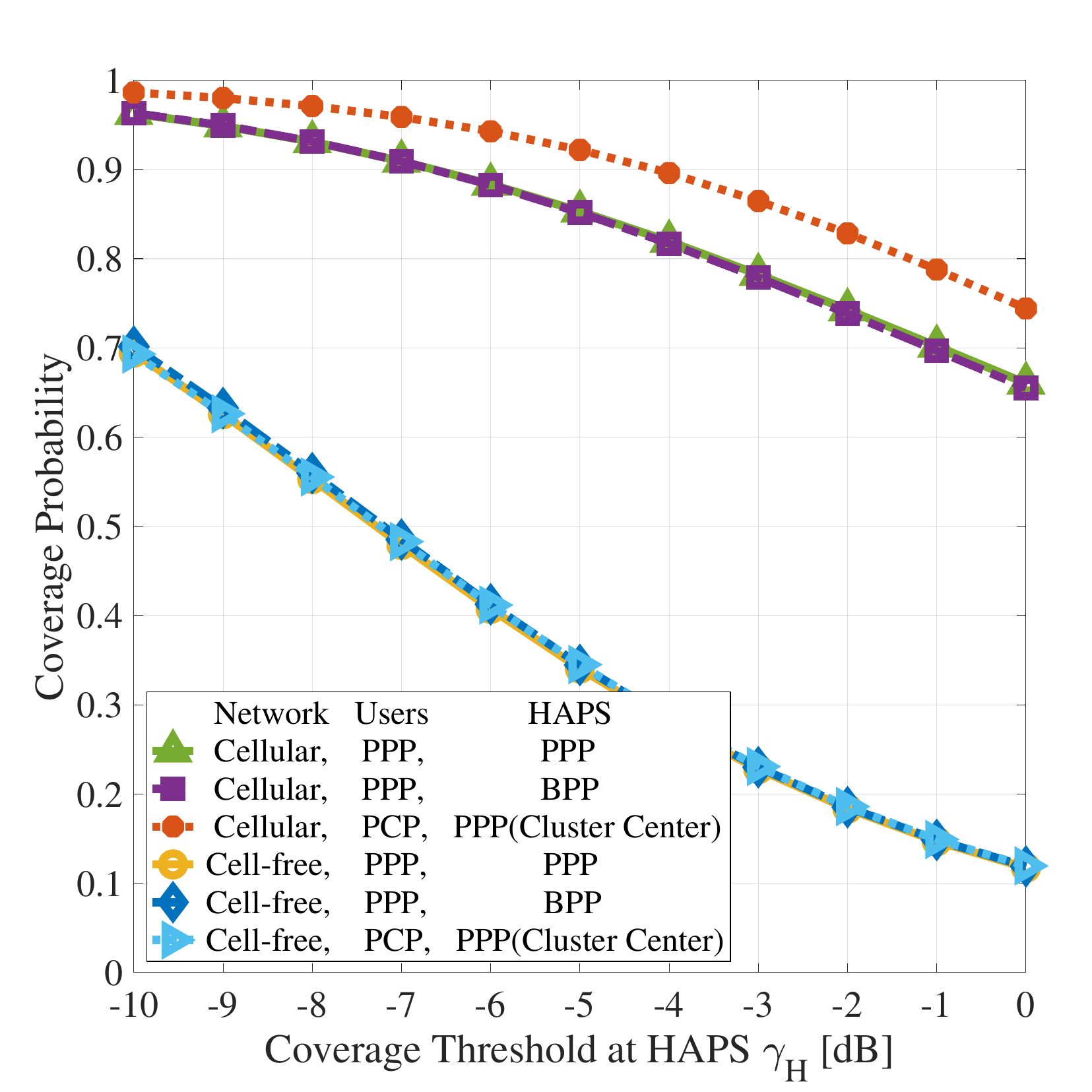}
\caption{Influence of stochastic model on uplink coverage probability.}
\label{fig9}
\end{minipage}
\hfill
\begin{minipage}[t]{0.328\linewidth}
 \centering
\includegraphics[width=0.99\linewidth]{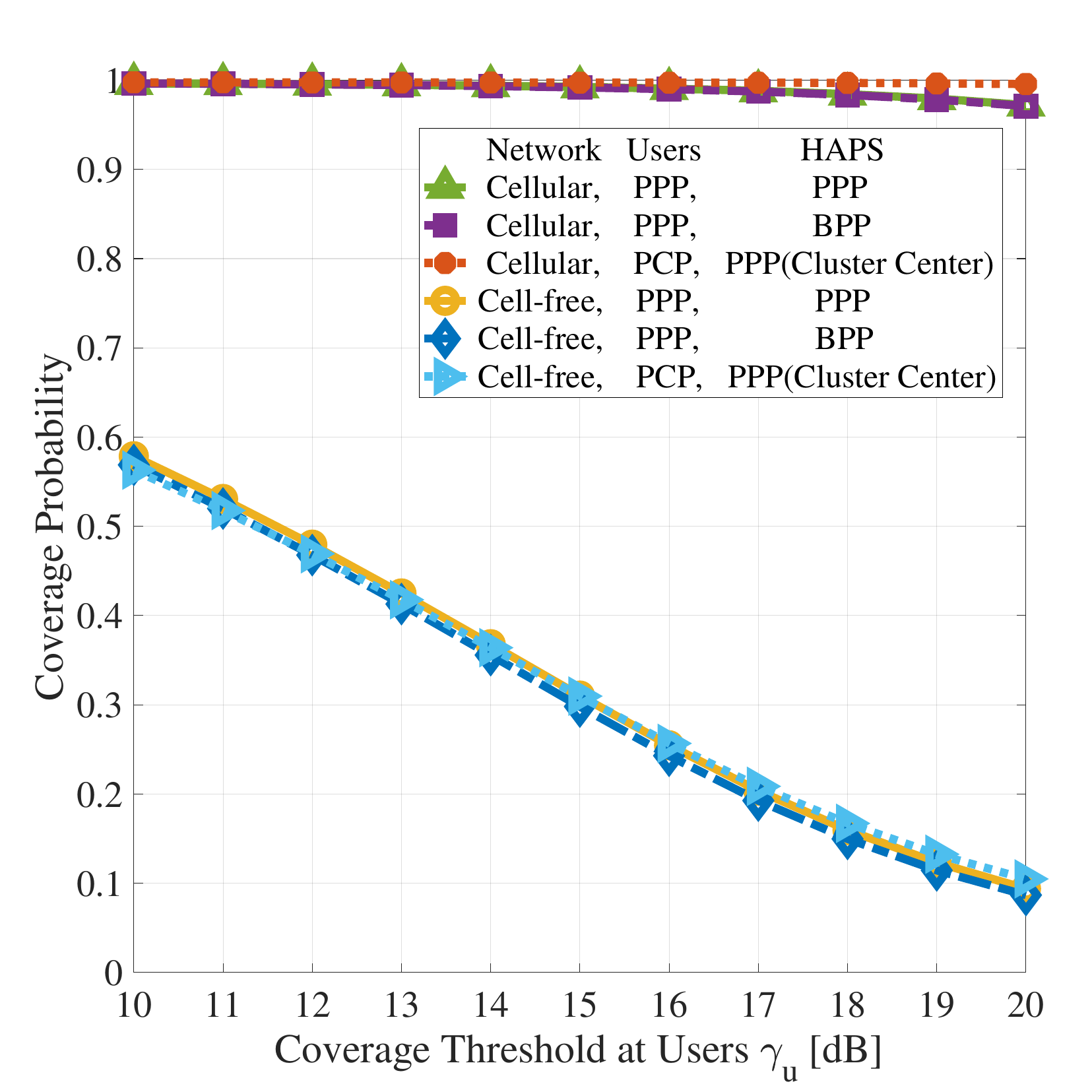}
\caption{Influence of stochastic model on downlink coverage probability.} 
\label{fig10}
\end{minipage}
\vspace{-0.4cm}
\end{figure*}

\vspace{-0.2cm}

 \subsection{Summary}
In what follows, we summarize the main takeaway insights. 
\begin{itemize}
    \item Although cell-free networks offer users more options, enhance network availability, and reduce interference, contact distance plays a more critical role in improving HAPS network coverage probability, ultimately leading to cellular networks tending to outperform cell-free networks.
    \item Employing directional antennas for transmitters and receivers can significantly enhance coverage probability. This improvement stems from the combined effect of interference reduction and focused signal energy, showcasing the potential for more efficient spectrum use in HAPS networks. 
    \item Increasing antenna gain initially improves coverage probability, but eventually levels off as signal and interference strengths rise. This insight suggests an optimal antenna gain threshold, beyond which additional gains do not translate into improved coverage probability.
    \item In cell-free networks, there is a counterintuitive negative correlation between HAPS density and downlink coverage probability, driven by the randomness of beam alignment. 
\end{itemize}


\section{Insights on Potential Open Issues and Research Directions} \label{IPOIRD}

This section explores several factors that align with the analytical framework of this article. The combination of these factors with the existing framework is a potential and interesting future research direction. 
\vspace{-0.2cm}
\subsection{Multi-Beam} 
\label{multi-beam}

To provide communication services for more users simultaneously, it is necessary to introduce multi-beam technology \cite{sellathurai2016user}, especially for communication devices with large coverage such as HAPS. In Fig.~\ref{fig7}, we consider that one HAPS is configured with the ability to transmit four beams with the Gaussian antenna pattern simultaneously in each sub-band. Fig.~\ref{fig7} shows that the introduction of a multi-beam communication mechanism has a little effect on downlink coverage probability. Because multi-beam communication will lead to a slight increase in downlink interference compared to single-beam communication. However, the cost can be justified since there is a substantial four times increase of the original channel capacity. 

In summary, the adoption of multi-beam technology, despite the slight increase in downlink interference, offers a significant boost in channel capacity, making it a valuable investment for enhancing communication services, particularly in large coverage scenarios such as HAPS.


\par

\vspace{-0.1cm}
\subsection{Other Models}
In addition to the PPP model adopted in this article, other stochastic point processes for HAPS and user modeling are also worth discussing.  As shown in Fig.~\ref{fig9} and Fig.~\ref{fig10}, modeling with PPP and BPP has a negligible effect on the coverage probability.

\par
In large-scale networks, another widely studied model for users who initiate service requests is the Poisson cluster process (PCP) \cite{andrews2011tractable}. In the PCP model, cluster centers are randomly distributed on the Earth's surface. For instance, a cluster center can be a city center or a gateway connecting the core network. Users follow a specific random distribution around these cluster centers. We assume cluster centers follow a spherical PPP with density $\lambda_H$ and radius $R_u$. HAPS are located directly above cluster centers with height $R_H-R_u$. The users to be served are distributed within a range of $50$~km from cluster centers. Users around each cluster center follow an independent PPP with a density of $1.3$~km$^2$. Overall, the users initiating the service follow a PCP and the average number of users is equal to users in a PPP with  density $\lambda_u$. Fig.~\ref{fig9} and Fig.~\ref{fig10} indicate that there is little difference in the downlink coverage performance between the PCP and PPP models, while the uplink coverage performance is slightly better under the PCP model than the PPP model. In general, different distribution models for users and HAPS are not the main factors influencing the coverage probability.

\vspace{-0.1cm}
\subsection{Limitation and Future Direction}

 SG framework is primarily applied in network performance evaluation. 
Specifically, performance evaluation aims to provide rapid and accurate assessments, and offer insight into system metrics under various parameters.
SG focuses on computational complexity and evaluation accuracy, rather than on proposing better strategies. Therefore, while the SG framework outperforms Monte Carlo simulation in terms of computational complexity, it does not lead to better system performance. This represents the inherent limitation of performance evaluation studies. 
An improved approach is to use SG framework to evaluate the performance of systems that integrate cutting-edge communication technologies, such as multiple-input multiple-output (MIMO) \cite{li2025downlink} and reconfigurable intelligent surface (RIS) \cite{shaikh2024performance}.

 In HAPS networks, multi-beam technology has proven effective in enhancing system coverage probability and throughput, as shown in Subsection IV-A. Combining multi-beam technology with MIMO arrays can further improve system performance \cite{goto2018leo}. Following the completion of this work, the downlink performance of MIMO satellite networks was investigated within the framework of spherical SG \cite{li2025downlink}. However, for simplicity, both transmitters and receivers were equipped with single antennas. The results in \cite{li2025downlink} can serve as a baseline for coverage analysis in multi-antenna models, which is left for future work. Based on this study, we believe that integrating MIMO technology, modeling channel matrices and antenna correlations, will open up opportunities for further advancing HAPS networks.

\vspace{-0.1cm} 
\section{Conclusion} \label{C}

This paper presented a tractable approach for evaluating performance in massive HAPS networks with the aid of the spherical SG framework. Accordingly, we derived analytical expressions of uplink and downlink coverage probability with directional transmitting and receiving beams in cellular and cell-free HAPS networks. Analytical results align with simulation results, thereby validating the accuracy of our model. We observe that cellular networks outperform cell-free networks, underscoring the importance of contact distance in HAPS network design. In addition, increasing antenna gain and reducing the half-power beamwidth demonstrate diminishing returns in terms of coverage improvement; accordingly, our analytical model can guide practitioners in balancing cost and performance. Moreover, this work can be extended to study the integration of multi-beam and MIMO technologies in HAPS networks, providing system-level insights.

\appendices

\section{Proof of Lemma~\ref{association}}\label{app:association}

According to the strongest average received power association strategy, a sufficient and necessary condition for the user associating with the typical HAPS is that there are no HAPS closer to the user than the typical HAPS. We use $\mathcal{S}_u(\theta)$ to present the spherical cap with a central angle as $2\theta$, whose rotation axis is the line connecting the center of the Earth and the user. Therefore, the probability of the user associating with the HAPS that has a polar angle difference of $\theta$ can be calculated as
\begin{IEEEeqnarray}{RCL}
     P^A_u \left( \theta \right) &=& \mathbb{P}\left[ {\mathcal{N}\left( {{\mathcal{S}_u(\theta)}} \right) = 0} \right] \notag  \\
    & \overset{(a)}{=} &\exp \left( -{\lambda}_H \mathcal{A}\left( \mathcal{S}_u(\theta) \right)  \right)  \frac{\exp \left(  - {\lambda}_H \mathcal{A}\left( \mathcal{S}_u(\theta) \right) \right)^0}{0!}  \IEEEeqnarraynumspace \notag  \\
    & =& \exp \left( -{\lambda}_H 2 \pi R_H^2 \left(1-\cos\theta\right) \right),
\IEEEeqnarraynumspace
\end{IEEEeqnarray}

$\mathcal{N}\left( \mathcal{S}_u(\theta) \right)$ counts the number of HAPS in the spherical cap $\mathcal{S}_u(\theta)$, and $\mathcal{A}\left( \mathcal{S}_u(\theta) \right)$ is the area measure of $\mathcal{S}_u(\theta)$. Step $(a)$ is from the general PPP \cite{primer}. Therefore, the probability of the user at $(R_u,\phi,\theta)$ associating with the typical HAPS at $(R_H,0,0)$ can be calculated as
\begin{equation}
    P_u^{A,{\rm{cll}}} \left( \theta \right) = \exp \left( - \widetilde{\lambda}_H 2 \pi R_H^2 \left(1-\cos\theta\right) \right),
\end{equation}
where $\widetilde{\lambda}_H$ is the density of HAPS in the user's LoS region after removing the typical HAPS,
\begin{equation}\label{lambdaH}
    \widetilde{\lambda}_H = \lambda_H - \frac{1}{2\pi R_H (R_H-R_u)}.
\end{equation}
Note that the association probability is only related to the polar angle $\theta$ of the user, not to the radius $R_u$ and azimuth angle $\phi$.

\section{Proof of Lemma~\ref{gain1}}\label{app:gain1}
According to Slivnyak's theorem \cite{feller1991introduction}, the distribution of homogeneous PPP is invariant with the rotation of the sphere. Furthermore, the association probability is not related to the azimuth angle. Therefore, the average antenna gain from a user with coordinate $(R_u,\phi,\theta)$ to the typical HAPS with coordinate $(R_H,0,0)$ is equal to the average antenna gain from a user with coordinate $(R_u,0,0)$ to the typical HAPS with coordinate $(R_H,0,\theta)$. Between them, the derivation of the latter is more concise. 
\par

Next, we want to derive the position the user's beam aligns with, i.e. the position of the associated HAPS of this user at $(R_u,0,0)$. The first case is when the user associates with the HAPS at $(R_H,0,\theta)$, which happens with probability $P^{A, \rm{cll}}_u \left(\theta\right)$. Therefore, the average gain, in this case, is,
\begin{equation}
    \overline{G}_{u,1}^{\rm{cll}} \left(\theta\right) = P^{A, \rm{cll}}_u \left(\theta\right) G_u(0).
\end{equation}

\par

The second case is the user associates with one of the other HAPS. Denote the associated HAPS' coordinate as $(R_H,\varphi,\psi)$. The corresponding beam deflection angle is $\mathcal{T}( R_H,0,\theta;$ $R_u,0,0; R_H,\varphi,\psi )$. The average antenna gain of the second case is,
\begin{IEEEeqnarray}{RCL}\label{appgain1}
    \overline{G}_{u,2}^{\rm{cll}} \left( \theta \right) &=& \mathbb{E}_{\varphi,\psi}[G_u\left( \mathcal{T}\left ( R_H,0,\theta;R_u,0,0;R_H,\varphi,\psi \right) \right)] 
 \notag \\
    & = & \int_0^{2\pi} \int_0^\theta G_u\left( \mathcal{T}\left ( R_H,0,\theta;R_u,0,0;R_H,\varphi,\psi \right) \right) \notag \\ 
    && \times f_\psi\left(\psi\right) f_\varphi\left(\varphi\right) \mathrm{d}\psi \mathrm{d}\varphi.
\end{IEEEeqnarray} 
From the invariance of the azimuth angle with respect to the association probability, $\varphi$ is uniformly distributed between $0$ and $2\pi$.
On the other hand, the probability density function (PDF) of $\psi$ is derived as follows,
\begin{IEEEeqnarray}{RCL}\label{appgain2}
     f_\psi\left(\psi\right) &=& \frac{\mathrm{d} F_\psi\left(\psi\right)}{\mathrm{d} \psi}  \notag 
 \\
    &=& \frac{\mathrm{d}}{\mathrm{d} \psi} \left( 1 - \exp \left( -\widetilde{\lambda}_H 2 \pi R_H^2 \left(1-\cos\psi\right) \right) \right) \notag  \\
    &=& \widetilde{\lambda}_H 2 \pi R_H^2 \sin\psi \exp \left( -\widetilde{\lambda}_H 2 \pi R_H^2 \left(\!1\!-\!\cos\psi\right) \right),
\IEEEeqnarraynumspace
\end{IEEEeqnarray}
where $0\leq \psi \leq \theta$ and $\widetilde{\lambda}_H$ is defined in (\ref{lambdaH}). Substitute (\ref{appgain2}) into (\ref{appgain1}), $\overline{G}_{u,2}^{\rm{cll}} \left( \theta \right)$ is derived. Note that $f_\psi\left(\psi\right)$ is called the PDF of contact angle distribution, and the contact angle is defined as the central angle between a user and the HAPS associated with the user \cite{al2021modeling}. Finally, combining the two cases, the final average antenna gain is $\overline{G}_u^{\rm{cll}}(\theta) = \overline{G}_{u,1}^{\rm{cll}}(\theta) + \overline{G}_{u,2}^{\rm{cll}}(\theta)$.

 \section{Proof of Lemma~\ref{LT1}}\label{app:LT1}
We assume that the typical HAPS at $(R_H,0,0)$ is serving a user at $x_0(R_u,\phi_0,\theta_0)$. Similarly, according to Slivnyak's theorem, the distribution of interfering users is independent of the position of the serving user. In addition, the association probability and average uplink antenna gain are only related to $\theta_0$. Therefore, the Laplace transform of total interfering power when the user is located at coordinate $x_0(R_u,\phi_0,\theta_0)$ is equivalent to that when the user is located at coordinate $x_0(R_u,0,\theta_0)$. Then, the Laplace transform of total interfering power from other users can be derived from its definition,

\begin{IEEEeqnarray}{RCL}\label{appc-1}
     && \mathcal{L}_u^{\rm{cll}}  (s,\theta_0) = \mathbb{E}_I [\exp (-sI (\theta_0)]  \notag  \\ 
    && \overset{(a)}{=}\mathbb{E}_{\mathcal{X},W_I} \left[ \exp \left(\sum_{x_i \in \mathcal{X} \backslash x_0} -s \rho^r_{u,x_i} (\theta_0) \right) \right] \notag  \\
    & &\overset{(b)}{=}  \!\mathbb{E}_{\mathcal{X}} \!\!\!\left[ \! \prod_{x_i \in \mathcal{X} \backslash x_0}\!\! \!\! \left( \!\! \frac{m}{m\!+\!s\rho_u^t G_u(\theta_i^t) 
    G_H(\theta_i^r) \left(\!\frac{\nu}{4\pi}\!\right)^2 \!\zeta D_i^{-2} } \right)^m \right],
\IEEEeqnarraynumspace
\end{IEEEeqnarray}
where the equation in step (a) is satisfied because of the independence of the users' positions and small-scale fading. Step (b) follows the moment-generating function (MGF) of Nakagami-$m$ fading \cite{annamalai2003moment}.

\par
Then, Step (c) in 
(\ref{appc-2}) follows the probability generating functional (PGFL) of the PPP \cite{primer}. We calculate the Laplace transform of the total interference power by traversing all interfering users within the LoS region of the typical HAPS. The LoS region is divided into area elements 
$\mathcal{S}_{\odot}(\varphi,\varphi+\mathrm{d}\varphi; \psi,\psi+\mathrm{d}\psi)$ with a fixed radius, whose azimuth angle varies from $\varphi$ to $\varphi+\mathrm{d}\varphi$, and polar angle varies from $\psi$ to $\psi+\mathrm{d}\psi$ (we denote the area element as $\mathcal{S}_{\odot}$ for convenient). The average uplink antenna gain $\overline{G}_u^{\rm{cll}}(\psi)$ at the typical HAPS from users in area element $\mathcal{S}_{\odot}$ is defined in (\ref{gainccl}). Given the position of serving user $(R_u,0,\theta_0)$, the average downlink antenna gain at users in area element $\mathcal{S}_{\odot}$ from the typical HAPS is $G_H(\mathcal{T}\left( R_u,0,\theta_0;R_H,0,0;R_u,\varphi,\psi \right) )$. The Euclidean distance between the typical HAPS to the interfering  $d(R_H,0,0;R_u,\varphi,\psi)$ is defined in (\ref{Euclidean}). Finally, $\mathcal{N}(\mathcal{S}_{\odot})$ counts the number of interfering users in area element $\mathcal{S}_{\odot}$. Through the above analysis, the results in (\ref{appc-1}) can be derived as,
\vspace{-0.2cm}
\begin{IEEEeqnarray}{RCL}\label{appc-2}
    & \!\mathbb{E}_{\mathcal{X}}& \left[ \! \prod_{x_i \in \mathcal{X} \backslash x_0}\!\! \!\! \left( \!\! \frac{m}{m\!+\!s\rho_u^t G_u(\theta_i^t) 
    G_H(\theta_i^r) \left(\!\frac{\nu}{4\pi}\!\right)^2 \!\zeta D_i^{-2} } \right)^m \right] \notag  \IEEEeqnarraynumspace  \\
    & \overset{(c)}{=}& \int_{\varphi} \int_{\psi} m^m \cdot \Big( m+s\rho_u^t \overline{G}_u^{\rm{cll}}(\psi) \notag  \\
    &&\times G_H(\mathcal{T}\left( R_u,0,\theta_0;R_H,0,0;R_u,\varphi,\psi \right) )  \notag \\
    && \times \left(\frac{\nu}{4\pi}\right)^2 \zeta \, \left(d(R_H,0,0;R_u,\varphi,\psi)\right)^{-2} \Big)^{-m} \notag  \\
    &&\times \mathcal{N}(\mathcal{S}_{\odot}(\varphi,\varphi+\mathrm{d}\varphi; \psi,\psi+\mathrm{d}\psi)) \notag  \\
    & \overset{(d)}{=}& \int_0^{2\pi} \int_0^{\Theta} m^m \cdot \bigg( m+s\rho_u^t \overline{G}_u^{\rm{cll}}(\psi)  \notag  \\
    &&\times G_H(\mathcal{T}\left( R_u,0,\theta_0;R_H,0,0;R_u,\varphi,\psi \right) ) \notag 
  \\
    &&\times \left(\frac{\nu}{4\pi}\right)^2 \zeta \left(d(R_H,0,0;R_u,\varphi,\psi)\right)^{-2} \bigg)^{-m}  \notag 
 \\
    &&\times \left( \lambda_u - \frac{{R_H}}{2\pi R_u^2 ({R_H}-{R_u})}\right) R_u^2 \sin\psi \mathrm{d}\varphi \mathrm{d}\psi,
\end{IEEEeqnarray}
where $\Theta=\arccos({R_u}/{R_H})$ is the maximum polar angle of the typical HAPS' LoS region. 
In step (d), $\mathcal{N}(\mathcal{S}_{\odot})=\widetilde{\lambda}_u \mathcal{A}(\mathcal{S}_{\odot})$, where 
$\widetilde{\lambda}_u$ is the average density of user in $\mathcal{S}_{\odot}$ excluding the serving user,
\begin{equation}\label{tildelambdau}
    \widetilde{\lambda}_u = \lambda_u \frac{\lambda_u 2\pi R_u^2 (1-\cos\Theta) - 1}{\lambda_u 2\pi R_u^2 (1-\cos\Theta)},
\end{equation}
and $\mathcal{A}(\mathcal{S}_{\odot})=R_u^2 \sin\psi \mathrm{d}\varphi \mathrm{d}\psi$ is the area measure of $\mathcal{S}_{\odot}$.

\vspace{-0.1cm}
\section{Proof of Theorem~\ref{cov1}}\label{app:cov1}
Considering the bidirectional beam alignment between typical HAPS and service users being achieved, signal receiving power at the typical HAPS is,
\begin{equation}
    \rho_H^r = \rho_u^t G_u(0) G_H(0) \left(\frac{\nu}{4\pi}\right)^2 \zeta D_S^{-2} W_S,
\end{equation}
where $D_S$ is a random variable called contact distance that records the distance between the typical HAPS and its serving user \cite{talgat2020nearest}. The relationship between the PDF of contact distance and the PDF of the contact angle is,
\begin{IEEEeqnarray}{RCL}\label{contactangle}
    f_{\theta_0} (\theta) & =& {\lambda}_H 2 \pi R_H^2 \sin\theta \exp \left( -{\lambda}_H 2 \pi R_H^2 \left(1-\cos\theta\right) \right) \notag \\ 
    & =& f_{D_s}\left( \sqrt{R_H^2 + R_u^2 -2 R_H R_u \cos\theta} \right).
\IEEEeqnarraynumspace
\end{IEEEeqnarray} 

\par
Next, we derive the analytical expression for the coverage probability from the definition,
\begin{IEEEeqnarray}{RCL}\label{appd-1} 
     P_u^{C,{\rm{cll}}}  (\gamma_H)& =& \mathbb{P} \left[ \frac{\rho_H^r}{I +\sigma^2} > \gamma_H \right] \notag  \\
    & =& \mathbb{P} \left[ W_S > \frac{\gamma_H \cdot (I+\sigma^2)}{\rho_u^t G_u(0) G_H(0) \left(\frac{\nu}{4\pi}\right)^2 \zeta D_S^{-2}} \right] \notag  \\
    & =& \int_0^{\Theta} \mathbb{E}_{I} \Bigg[ \overline{F}_{W_S} \Bigg( \frac{ \left(R_H^2 + R_u^2 -2 R_H R_u \cos\theta\right) }{\rho_u^t G_u(0) G_H(0) \left(\frac{\nu}{4\pi}\right)^2 \zeta } \notag  \\
    && \times\gamma_H  (I+\sigma^2) \Bigg) \Bigg]  {\lambda}_H 2 \pi R_H^2 \sin\theta \notag  \\
    && \times \exp \left( -{\lambda}_H 2 \pi R_H^2 \left(1-\cos\theta\right) \right) \mathrm{d}\theta, 
\end{IEEEeqnarray}
where $\Theta=\arccos{\frac{R_u}{R_H}}$ is the upper bound of polar angle for the typical HAPS' LoS region, and $\overline{F}_{W_S} (\cdot)$ is the complementary cumulative distribution function (CCDF) of the shadowed Rician fading.

\par
We take the well-known approximation of the CDF of the incomplete Gamma distribution \cite{alzenad2019coverage}, the result of the approximation is proved to provide a tight upper bound for the incomplete Gamma distribution \cite{bai2014coverage},
\begin{equation}
\frac{\Gamma_l(m,mg)}{\Gamma(m)} \lesssim \left( 1-\exp\left(- \left(m!\right)^{\frac{-1}{m}} mg \right) \right)^m.
\end{equation}
Applying the above approximation, the CCDF of the shadowed Rician fading can be further written as,
\begin{IEEEeqnarray}{RCL}\label{appd-2}
     \mathbb{E}_{I} && \left[\overline{F}_{W_S} \left( \frac{\gamma_H  (I+\sigma^2)  \left(R_H^2 \!+ \!R_u^2 \!-\!2 R_H R_u \cos\theta\right) }{\rho_u^t G_u(0) G_H(0) \left(\frac{\nu}{4\pi}\right)^2 \zeta } \right) \right]  \notag  \\
     \gtrsim && 1 - \left( \frac{2b_0 n}{2 b_0 n + \Omega} \right)^n \sum_{z=0}^{\infty} \frac{(n)_z}{z!} \left( \frac{\Omega}{2 b_0 n + \Omega} \right)^z \notag  \\
    && \times \mathbb{E}_I \Bigg[ \Bigg( 1 -  \exp \Bigg( -((z+1)!)^{\frac{-1}{z+1}} \gamma_H  (I+\sigma^2)  \notag  \\
    && \times \frac{\left(R_H^2 + R_u^2 -2 R_H R_u \cos\theta\right) }{2b_0 \rho_u^t G_u(0) G_H(0) \left(\frac{\nu}{4\pi}\right)^2 \zeta } \Bigg) \Bigg)^{z+1} \Bigg].
), \IEEEeqnarraynumspace
\end{IEEEeqnarray} 
Define $\beta_u(\theta,\gamma_H)$ as
\begin{IEEEeqnarray}{RCL}
        \beta_u(\theta,\gamma_H) =&& ((z+1)!)^{\frac{-1}{z+1}} \notag \\ 
        && \times \frac{\gamma_H \cdot \left(R_H^2 + R_u^2 -2 R_H R_u \cos\theta\right) }{2b_0 \rho_u^t G_u(0) G_H(0) \left(\frac{\nu}{4\pi}\right)^2 \zeta },
\IEEEeqnarraynumspace
\end{IEEEeqnarray} 
then we have,
\begin{IEEEeqnarray}{RCL}\label{appd-3}
     \!\!\mathbb{E}_I && \left[ \left( 1\! -\!  \exp \left( \!-((z+1)!)^{\frac{-1}{z+1}}  (I+\sigma^2) \beta_u(\theta,\gamma_H)\! \right) \! \!\right)^{z+1} \!\right]  \notag  \\
    \overset{(a)}{=} && 1 - \mathbb{E}_I \Bigg[ \sum_{k=1}^{z+1} \binom{z+1}{k} (-1)^{k+1} \notag  \\
    &&  \times\exp\big(-k\beta_u(\theta,\gamma_H) \left(I+\sigma^2\right) \big) \Bigg] \notag \\
     = && 1 - \sum_{k=1}^{z+1} \binom{z+1}{k} (-1)^{k+1} \notag  \\
    &&  \times \exp\left(-k\beta_u(\theta,\gamma_H) \sigma^2 \right) \mathcal{L}_u^{\rm{cll}}(k\beta_u(\theta,\gamma_H),\theta),
\IEEEeqnarraynumspace
\end{IEEEeqnarray}
where step (a) follows from the binomial theorem, and $\mathcal{L}_u^{\rm{cll}}(k\beta_u(\theta,\gamma_H),\theta)$ is defined in Lemma~\ref{LT1}. Finally, the proof is finished by substituting (\ref{appd-2}) and (\ref{appd-3}) into (\ref{appd-1}).

\section{Proof of Lemma~\ref{gain2}}\label{app:gain2}
In cell-free networks, each HAPS within the user's LoS region is associated with equal probability. Since the average number of HAPS within the LoS region is $\lambda_H 2 \pi R_H \left( R_H - R_u \right)$. Therefore, the probability that a user in the typical HAPS' LoS region is associated with this HAPS is given as,
\begin{equation}\label{assocfup}
    P_u^{A,{\rm{cf}}} = \left(\lambda_H 2 \pi R_H \left( R_H - R_u \right) \right)^{-1}.
\end{equation}
\par
In addition, the user's associated HAPS cannot be further away from the user than the typical HAPS in cellular networks. However, cell-free networks do not have this limitation, so the integral interval of $\psi$ needs to be transformed accordingly. The remaining proof of Lemma~\ref{gain2} is similar to that of Lemma~\ref{gain1}, and the association probability can be written as,
\begin{IEEEeqnarray}{RCL}\label{appE-2}
    \!\!\!\!\overline{G}_u^{\rm{cf}} &&  \left( \theta \right) \! = \! P_u^{A,{\rm{cf}}} G_u(0) \!+ \!\int_0^{2\pi} \!\!\! \int_0^{\Theta} \!\! \lambda_H \left( 1 \!- \!P_u^{A,{\rm{cf}}} \right) R_H^2  \sin\psi  \notag  \IEEEeqnarraynumspace \\
    && \!\!\!\!\!\! \times    P_u^{A,{\rm{cf}}}   G_u\left( \mathcal{T}\left (R_H,0,\theta; R_u,0,0; R_H,\varphi,\psi \right) \right) \mathrm{d}\psi \mathrm{d}\varphi.
\end{IEEEeqnarray} 
Substitute (\ref{assocfup}) into (\ref{appE-2}), and Lemma~\ref{gain2} is proved.

\section{Proof of Lemma~\ref{gain3}}\label{app:gain3}
As the same with the proof in Lemma~\ref{gain1}, according to Slivnyak's theorem and the rotation invariance of the PPP distribution, we consider analyzing the average antenna gain from the HAPS with coordinate $(R_H,0,0)$ to the typical user with coordinate $(R_u,0,\theta)$. 

\par
The total number of users associated with the HAPS in area element $\mathcal{S}_{\odot}(\varphi,\varphi+\mathrm{d}\varphi; \psi,\psi+\mathrm{d}\psi)$ is $\widetilde{\lambda}_u R_u^2 \sin\psi P_u^{A,\rm{cll}}(\psi) \mathrm{d}\varphi \mathrm{d}\psi$, where $\widetilde{\lambda}_u$ is defined in (\ref{tildelambdau}), and $P_u^{A,\rm{cll}}(\psi)$ is defined in Lemma~\ref{association}. The area element $\mathcal{S}_{\odot}$ is defined in Appendix~\ref{app:LT1}. Assuming that a user is associated with the HAPS at $(R_H,0,0)$, the probability of the user locating in $\mathcal{S}_{\odot}$ is equal to the ratio of the number of users in $\mathcal{S}_{\odot}$ to the number of all users in HAPS' LoS region. The latter can be rewritten as,
\begin{IEEEeqnarray}{RCL}
       && \frac{\widetilde{\lambda}_u R_u^2  \sin\psi P_u^{A,\rm{cll}}(\psi) \mathrm{d}\varphi \mathrm{d}\psi}{\int_0^{2\pi} \int_0^{\Theta} \widetilde{\lambda}_u R_u^2 \sin\psi P_u^{A,\rm{cll}}(\psi) \mathrm{d}\varphi \mathrm{d}\psi} 
       \notag  \\ 
       && \,\, = \frac{ \sin\psi P_u^{A,\rm{cll}}(\psi) \mathrm{d}\psi \mathrm{d}\varphi}{2\pi \int_0^{\Theta} \sin\psi P_u^{A,\rm{cll}}(\psi) \mathrm{d}\psi},
\end{IEEEeqnarray}
where $\Theta=\arccos({R_u}/{R_H})$. When the HAPS at $(R_H,0,0)$ is serving with a user in area element $\mathcal{S}_{\odot}$, the corresponding beam deflection angle between the typical user and the serving user is $\mathcal{T}(R_u,0,\theta;R_H,0,0;R_u,\varphi,\psi)$. The average antenna gain is,
\begin{IEEEeqnarray}{RCL}
    \overline{G}_H^{\rm{cll}} \left( \theta \right) &= &\mathbb{E}_{\varphi,\psi}[G_H\left( \mathcal{T}\left( R_u,0,\theta;R_H,0,0;R_u,\varphi,\psi \right) \right)] \notag  \\
    &=& \int_0^{2\pi} \int_0^\Theta G_H\left( \mathcal{T}\left ( R_u,0,\theta;R_H,0,0;R_u,\varphi,\psi \right) \right)  \notag  \\
    && \times \frac{ \sin\psi P_u^A(\psi) \mathrm{d}\psi \mathrm{d}\varphi}{2\pi \int_0^{\Theta} \sin\psi P_u^A(\psi) \mathrm{d}\psi}.
\IEEEeqnarraynumspace
\end{IEEEeqnarray}

\par
Finally, note that the derived expression is an approximation of the average antenna gain. In the last step, we integrate the entire LoS region of the interfering HAPS. However, in the cellular network, users in the neighborhood of the typical user are most likely associated with the HAPS which is providing service to the typical user, rather than interfering HAPS. Therefore, the estimated average antenna gain from interfering HAPS will be slightly larger than the real value, resulting in a smaller value of estimated coverage probability. In the section on numerical simulation, it is proved that the estimated coverage probability provides a tight lower bound for the coverage probability obtained by Monte Carlo simulation.

\bibliographystyle{IEEEtran}
\bibliography{references}

\end{document}